\documentclass[
aps,%
11pt,%
final,%
titlepage,%
oneside,%
twocolumn,%
nobibnotes,%
nofootinbib,%
superscriptaddress,%
showkeys,%
centertags]%
{revtex4}

\usepackage{amsmath}
\usepackage{graphicx}
\usepackage[colorlinks=true,linkcolor=blue,citecolor=blue,urlcolor=blue]{hyperref}

\usepackage{tabularx}

\newcommand{\ab}{Astroph. Bull.}
\newcommand{\astrep}{Astron. Rep.}

\newcommand{\aap}{A\&A} 
\newcommand{\afz}{Afz} 
\newcommand{\aj}{AJ} 
\newcommand{\apjl}{ApJ}
\newcommand{\apjs}{ApJS}
\newcommand{\mnras}{MNRAS} 
\usepackage{amssymb}

\setcitestyle{authoryear}

\usepackage{setspace}

\begin{document}

\singlespacing

\title{\LARGE The line-of-sight analysis of spatial distribution of galaxies in the COSMOS2015 catalogue}



\author{\firstname{Maxim}~\surname{Nikonov}}
\email{maxnikonov007@gmail.com}
\affiliation{The Moscow Institute of Physics and Technology, 1 ``A'' Kerchenskaya St., Moscow, 117303, Russia}

\author{\firstname{Mikhail}~\surname{Chekal}}
\email{mishka.chekal@gmail.com}
\affiliation{The Moscow Institute of Physics and Technology, 1 ``A'' Kerchenskaya St., Moscow, 117303, Russia}

\author{\firstname{Stanislav}~\surname{Shirokov}}
\email{arhath.sis@yandex.ru}
\affiliation{SPb Branch of Special Astrophysical Observatory of Russian Academy of Sciences, 65 Pulkovskoye Shosse, St Petersburg 196140, Russia}

\author{\firstname{Andrey}~\surname{Baryshev}}
\email{baryshev.andriy@gmail.com}
\affiliation{Saint Petersburg Electrotechnical University, Ulitsa Professora Popova 5, 197376 St. Petersburg, Russia}

\author{\firstname{Vladimir}~\surname{Gorokhov}}
\email{vlgorohov@mail.ru}
\affiliation{Saint Petersburg Electrotechnical University, Ulitsa Professora Popova 5, 197376 St. Petersburg, Russia}

\begin{abstract} 
New observations of high-redshift objects are crucial for the improvement of the standard $\Lambda$CDM cosmological model and our understanding of the Universe. One of the main directions of modern observational cosmology is the analysis of the large-scale structure of Universe, in particular, in deep fields. 
We study the large-scale structure of the Universe along the line of sight using the latest version of the COSMOS2015 catalogue, which contains 518,404 high quality photometric redshifts of galaxies selected in the optical range of the COSMOS field ($2\times 2$ deg$^2$), with depth up to the redshift $z \sim 6$. 
We analyze large-scale fluctuations in the number of galaxies along the line of sight and provide an estimate of the average linear sizes of the self-correlating fluctuations (structures) in independent redshift bins of $ \Delta z = 0.1 $ along with the estimate of the standard deviation from homogeneity (the observed cosmic variance).
We suggest a new method of the line-of-sight analysis based on previous works and formulate further prospects of method development.
For the case of the theoretical form of approximation of homogeneity in the $\Lambda$CDM framework, the average standard deviation of detected structures from homogeneity is $ \sigma_\text{mean}^{\Lambda \text{CDM}} = 0.09 \pm 0.02 $, and the average characteristic size of structures is $ R_\text{mean}^{\Lambda \text{CDM}} = 790 \pm 150 $ Mpc.
For the case of the empirical approximation of homogeneity, the average standard deviation of detected structures from homogeneity is $ \sigma_\text{mean}^\text{empiric} = 0.08 \pm 0.01 $, and the average characteristic size of structures is $ R_\text{mean}^\text{empiric} = 640 \pm 140 $ Mpc. 
\end{abstract}

\newcommand{\orcidauthorA}{0000-0001-5475-6151} 
\newcommand{\orcidauthorB}{0000-0002-2085-9045} 
\newcommand{\orcidauthorC}{0000-0002-8386-7659} 
\newcommand{\orcidauthorD}{0000-0003-4732-5729} 
\newcommand{\orcidauthorE}{0000-0002-8206-9890} 

\maketitle

\section{Introduction}
    \label{sec:Introduction}
    
The development of the observational techniques and increase of computing power in the beginning of the 21st century made it possible to study the evolution of large-scale structure of the Universe (LSSU) from the moment of birth of the first galaxies to modern era. The relevance of theoretical models and the correct understanding of the evolution of the Universe is determined by using various observational cosmological tests~\citep{Peebles1993,Lukash2010,Baryshev2012}.

The $\Lambda$CDM model is the standard cosmological model (SCM) that assumes homogeneous distribution of matter in the Universe (``at sufficiently large scales'')
including cold dark matter and dark energy.
The SCM also implies the evolution of the density fluctuations of both dark and luminous  matter with time, which is associated with the observed large-scale structure of the Universe.
The SCM predicts that the primordial small density fluctuations of the dark and luminous matter ($\delta \rho / \rho \sim 10^{-5}$) have linear time-growth for  the structures having scales larger $\sim$ 10 Mpc at present epoch ~\citep{Lukash2010,Moster2011}. The largest predicted by $\Lambda$CDM structures have sizes about the scale of the 
Baryon Acoustic Oscillations $R_{bao} \sim 100$ Mpc
\citep{Knebe2018}.

The very important observational test of the SCM is to estimate the maximum amplitudes and sizes of largest structures visible in the high redshift Universe. This test can establish an observational limit to ``the galaxy bias factor'', i.e. the ratio of fluctuation amplitudes of visible to dark matter ~\citep{Moster2011}.
Modern cosmological N-body simulations
\footnote{\url{https://www.cosmosim.org}} provide   
MULTIDARK-GALAXIES  catalogues,
derived from the Planck cosmology MULTIDARK simulations MDPL2,
with a volume of $1\,h^{-1}$Gpc$^{3}$
and mass resolution of 
$1.5\times 10^{9} h^{-1}M_{\odot}$
by applying the
semi-analytic models GALACTICUS, SAG, and SAGE
\citep{Knebe2018}.
These catalogues can be used for comparison between $\Lambda$CDM model predictions with real galaxy distribution at high redshift.
In the next  decade, the sensitivity limit of the upcoming telescopes
(The James Webb Space Telescope\footnote{\url{https://www.jwst.nasa.gov}}, ALMA\footnote{\url{https://www.almaobservatory.org}}, SKA\footnote{\url{https://www.skatelescope.org}}) will be sufficient to detect the earliest galaxies and hence to perform the ``largest structure'' observational test. 

However, already now we can estimate the amplitudes and sizes of visible matter fluctuations
by using the deepest  narrow-angle catalogue COSMOS2015~\citep{Laigle2016}, which contains more than $5\times10^5$ galaxies with high quality measured photometric redshifts up to 
$z=6$.

The COSMOS2015 catalogue is significantly improved compared with its previous version, used in the works by~\citet{Nabokov2010a, Nabokov2010b, Shirokov2016}. 
Note that in recent works, several structures were detected in the COSMOS field,
which also point to existence of large filamentary structures at $z\sim0.73$, called the COSMOS wall~\citep{Iovino2016}, structures at redshifts of $ 0.1 < z < 1.2 $~\citep{Darvish2017}, voids at $z\sim2.3$ ~\citep{Krolewski2018} and massive proto-supercluster at $z\sim2.45$~\citep{Cucciati2018}. The very large structure of the dark matter with size about 1,000 Mpc was detected in the COSMOS field by using method of the weak gravitational lensing 
\cite{Massey2007b, Massey2007}.

In this paper, we develop preceding approach used
by~\citet{Nabokov2010a, Nabokov2010b, Shirokov2016}. We develop a new method of analysis of the galaxy number counts by introducing two level of fluctuation sequences. We also use the last version of the COSMOS2015 catalogues.


\section{The COSMOS2015 catalogue} 
    \label{sec:Observational Data}

\subsection{Description}

In this paper, we use the COSMOS2015 catalogue~\citep{Laigle2016} of precise photometric redshifts to analyze the line-of-sight distribution of galaxies. The catalogue contains 30-band photometry data over the entire spectral range from radio to X-ray for 518,404 photometric redshifts of galaxies in the range $0 < z < 6$. 
The targets were selected in the optical range using the Hubble Space Telescope and were supplemented by data from the Chandra and XMM-Newton space observatories. 
The main goal of the COSMOS project is to study the relationships between the large-scale structure of Universe, dark matter, the formation of galaxies and the activity of nuclei in galaxies, as well as the influence of environmental conditions on the evolution of galaxies. The survey covers two square degrees on the celestial sphere nicknamed the COSMOS field in the direction of Sextans constellation. 

\subsection{Photometric redshifts}

Large catalogs of redshifts allow usage of a wide range of statistical tools that provide estimates of systematic effects via cosmological tests (e.g., the LSSU, Hubble diagram, Malmquist bias, and gravitational lensing bias~\citep{Scolnic2019,Shirokov2020}).
Conclusions about the LSSU drawn from careful analysis of photometric redshifts are mostly in agreement with results obtained from spectral surveys and other independent studies (see, for example,~\citet{Shirokov2016, SokolovJr2018}).

However, relatively large errors in the photometric redshifts estimation make the study of the LSSU especially challenging. 
Essential improvements in photometric redshift techniques, using the best SED fitting in the COSMOS field imaged in a large number of filters~\citep{Laigle2016} or deep learning methods~\citep{Pasquet2019}, allow to reach a redshift uncertainty
$\sigma_z = 0.007 (1+z)$ at small redshifts, which corresponds to a distance uncertainty of $\sim$ 40 Mpc (at $z\sim1$). 
At high redshifts $3<z<6$, the photometric redshifts in COSMOS2015 catalogue are defined on the grid in steps of
$\sigma_z<0.021(1+z)$~\citep{Laigle2016} using \emph{Hyperz: Photometric Redshift Code}\footnote{\url{https://www.researchgate.net/publication/258555988_Hyperz_Photometric_Redshift_Code}}.

In our analysis, we take the linear size of the redshift bin $\Delta z = 0.1$, which is an order of magnitude larger than the redshift error $\sigma_z \sim 0.007$. So, we can firmly study the large-scale structures having linear sizes larger than $\sim$ 100 Mpc.

\subsection{Selection effects}

In the literature there is no a reliable theoretical estimate of all observational selection effects in construction of the number-distance relation.
For example, the Malmquist bias is a result of the limited magnitude sensitivity of the equipment, which leads to preferable detection of the brighter sources. Such selection effects, as K-corrections, evolution effects, types of continuous spectra of galaxies, also must be taken into account in directly observed quantities.

Usual approach for taking into account the main selection effects in the primary observed number-distance relation is  to use a fitting function for the redshift distribution $N(z)$.
We use several  functional types of  combined observational selection effects, including an exponential function for all redshifts, which decreases sharply with increasing redshift. 

Photometric redshifts are determined by using simultaneously several filters, therefore, they are affected by the selection of visible magnitudes in different filters. This selection may contain the so-called ``spectral deserts'', which are often found in spectroscopic observations.
The description of this and other issues in detail, and the data processing decisions of the authors of the catalogue is given in~\citet[Sec.~3.2]{Laigle2016}.
In the COSMOS2015 catalogue, the observational selection effects (such as K-corrections, evolution effects, types of continuous spectra of galaxies and much more) have already been taken into account. Thus, we consider the catalogue as a fair galaxy sample.

\begin{figure*} \centering
    \includegraphics[width=\linewidth]{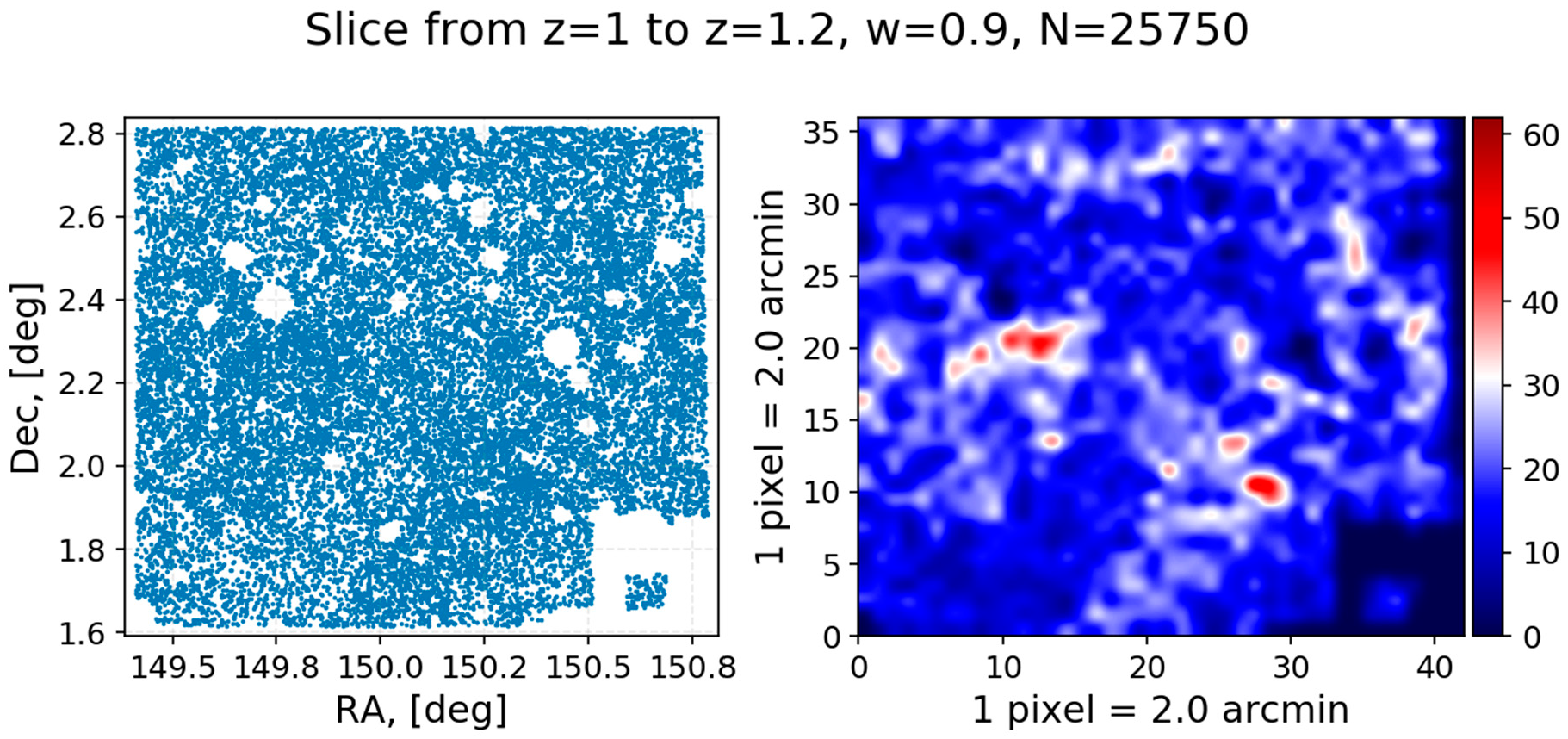}
    \caption{The projection on the sky of 25,750 galaxies from the COSMOS2015 catalogue in the range $1.0<z<1.2$ with the error $\sigma(z)<10$\%. The left angular map shows galaxies as points. The right angular map is gaussian interpolation in 36x42 pixels (each of which is equal to 2 squared minutes) plotted by using the \emph{matplotlib.pyplot} library. The color palette shows the galaxy count in one pixel. }
    \label{fig1:RaDec}
\end{figure*}

Figure~\ref{fig1:RaDec} shows the angular distribution of 25,750 galaxies in the COSMOS2015 catalogue in a slice with thickness of $\Delta z = 0.2$ at $ z = 1.0 $. 
Each imaged galaxy has 1$\sigma$ uncertainty $\sigma(z)<10$\%.
We can see the inhomogeneity of this distribution. The contrast between the medium pixels and the maximum pixels reaches a factor 3. The analysis of similar slices was performed in~\citet{Chiang2014}.
Circular empty regions are a result of masking stars. Stars in the field greatly complicates the construction of 3D maps of galaxies, but does not affect the radial distribution along the line of sight on large scales (in redshift bins). 

In the next section, we describe the method we used to analyze the LSSU by radial fluctuations of the number of galaxies along the line of sight. 


\section{The method for line-of-sight analysis of the LSSU in spatial bins}
    \label{sec:The Methods of Space Structure Analysis}

\subsection{Spatial distribution of galaxies}

According to the SCM, fluctuations in the large-scale distribution of galaxies could be approximated as a Gaussian process. 
Therefore, we assume that the histogram of the number of galaxies in spatial bins is described by a smooth function with a limited variance and an expected value (or average value) at each point. Moreover, this function starts to grow as $\sim R^D$, where $ R $ is the radius of probe sphere, and $ D $ is a certain density characteristic associated with the spatial correlation function of matter density. The quantity $D$ can be also called the fractal dimension, assuming that the large-scale distribution of matter is naturally-hierarchical one created by the gravity influence~\citep{Baryshev2012}).

According to the N-body simulations in the evolutionary theory of galaxies formation frameworks, the average number of galaxies has maximum at $z \sim 1 $, but their mass and luminosity are less than at $z=0$~\citep{Knebe2018}.
In other words, the detection probability of a massive and bright (detectable) galaxy is a bit lower at $z \sim 1 $ than at smaller redshifts. 
At large redshifts $ z > 2 $ the number of galaxies and their mass rapidly decrease.
This result can be verified by the query interface on CosmoSim website\footnote{\url{https://www.cosmosim.org}} by the MDPL2 catalogs~\citep{Knebe2018}.
Apparently, this is because small galaxies at $z \sim 1 $ are being continuously merged by the influence of gravity and other evolution processes, forming more massive ones at $z=0$. 
We also note that counts of simulated galaxies is a complicated issue and highly dependent on their detectability and detection methods. So, the work by~\citet{Graziani2017} shows how a smooth accretion can create MW-like galaxies at z=0 in overdense environment.

\subsection{Radial histogram of the number of galaxies}

Deep surveys of galaxies are narrow-angle conical sections of the global spatial distribution of galaxies. 
The radial distribution of the number of galaxies $N_1(z)$ is given by the equation
\begin{equation}
    \text{d} N(z,\text{d}z) = N_1(z) \text{d} z \, ,
    \label{eq:dN}
\end{equation}
where d$N$ is the number of galaxies in redshift range from $z$ to $z+\text{d}z$ (linear density). 
We consider the dependence $N(z)$ as the histogram $\Delta N/\Delta z$. The number of galaxies $\Delta N(z,\Delta z)$ counted in a spherical shell with thickness of $\Delta z$ such that
\begin{equation}
    \Delta N(z,\Delta z) = N(z) \Delta z \, .
    \label{eq:DeltaN}
\end{equation}
In other words, the $N(z)$ distribution is an observable approximation of the $N_1(z)$ distribution.

Thus, the distribution $ N (z) $ can be built in redshift bins with a step of $ \Delta z $. 
The value $ \Delta N (z, \Delta z) $ is the number of galaxies in a spherical shell $ (z, z + \Delta z) $. 
We conclude that with a reasonable choice of the parameter $ \Delta z $, the observed histogram $ \Delta N / \Delta z $ corresponds to the real radial distribution of the number of galaxies $ N (z) $. In our case, the reasonable choice of the parameter $ \Delta z $ is bin $\Delta z = 0.1$.

\subsection {The approximations of homogeneous distribution}

We approximate the homogeneous radial distribution of the number of galaxies with the least-squares method (LSM). 
Since the studied distribution has a high noise level (including the cosmic variance) and possible unknown systematic effects, the LSM via theoretical function may give an inaccurate result depending on the sample geometry. 
Therefore, the further introduced empirical functions as alternative can give a better approximation in the least-squares sense.  

In the SCM frameworks, the distribution of galaxies $N(z)$ is approximated by a power law~\citep{Ilbert2009}
\begin{equation}
    \Delta N_\text{model}(z,\Delta z) = A \left( \frac{z^a+z^{ab}}{z^b+C} \right) \Delta z \, ,
    \label{eq:N3(z)}
\end{equation}
where $N_\text{model} (z,\Delta z)$ is the number of galaxies in the redshift bin $(z, z+\Delta z)$. The formula has three free parameters for the LSM: $a$, $b$, and $C$.

We also used the empirical approximation function suggested in the works by~\citet{Massey2007,Lovyagin2009,Shirokov2016}
\begin{equation}
    \Delta N_\text{emp1}(z,\Delta z) = Az^\gamma e^{(-z/z_c)^\alpha}\Delta z \, ,
    \label{eq:N1(z)}
\end{equation}
where the free parameters $\gamma$, $\alpha$, and $z_c$ can be found by the LSM, and the dependent factor $A$ is the normalization constant. The normalization constant is chosen such that the integral of function $N(z)$ is equal to the total number of galaxies $N$,
\begin{equation}
    \int_0^\infty N(z) \text{d}z = \int_0^\infty 
    Az^\gamma exp \left( -\left(\frac{z}{z_c}\right)^\alpha \right) \text{d}z =
    \frac{Az_c^{\gamma+1}\Gamma(\frac{\gamma+1}{\alpha})}{\alpha} = N \, ,
    \label{eq:A}
\end{equation}
where $ \Gamma (x) $ is the Euler's complete gamma function. However, factor $ A $ cannot be calculated directly from the formula~(\ref{eq:A}) due to the high noise level. Therefore, accordingly to~\citet{Lovyagin2009}, it is necessary to search for it in the range from $ A-\sqrt{A} $ to $ A + \sqrt{A} $.  

In addition, we considered a particular case of the function~(\ref{eq:N1(z)}) for $ \alpha = 1 $ and new labels for the parameters,
\begin{equation}
    \Delta N_\text{emp2}(z,\Delta z) = Az^a e^{-bz}\Delta z \, ,
    \label{eq:N2(z)}
\end{equation}
to estimate the difference in the sum of LSM residuals between the three-parameter and two-parameter functions.

We searched for the coefficients using \emph{scipy} library and \emph{leastsq} function, which has two parameters: initial values of the parameters (input) and the sum of LSM residuals (output). A parametric vector ($10^4,\,1,\,1,\,1$), where the value $10^4$ corresponds to parameter $A$ in all approximation formulas~(\ref{eq:N3(z)}--\ref{eq:N2(z)}), was taken as a zero approximation for the parameters of all functions.
We call the approximation $\Delta N_\text{model}$ as the theoretical one and the approximations $\Delta N_\text{emp1}$ and $\Delta N_\text{emp2}$ as the empirical ones.
Since our analysis is based on the clear mathematical approach without an accurate physical interpretation of the approximation parameters, we will use quantities $\Delta N_\text{approx1}$, $\Delta N_\text{approx2}$, and $\Delta N_\text{approx3}$ instead of $\Delta N_\text{model}$, $\Delta N_\text{emp1}$, and $\Delta N_\text{emp2}$. 

\subsection{The fluctuation amplitudes}

After finding the best fitting parameters, we can detect inhomogeneities in the radial histogram of the observed number of galaxies. 
Fluctuations of physical quantities are the deviations of these quantities from their mean value, caused by random processes.
The variance of fluctuations can be found in the plot of relative fluctuations, which we call the fluctuation pattern. 
We consider a fluctuation in each redshift bin $\delta(z,\Delta z)$ as follows
\begin{equation}
    \delta(z,\Delta z) = (\Delta N_\text{obs}-\Delta N_\text{approx})/\Delta N_\text{approx} \, ,
\end{equation}
where $\Delta N_\text{obs}$ is the observed number of galaxies in the bin, and $\Delta N_\text{approx}$ is the theoretically expected number of galaxies in the bin that is equal to the integral under the approximation function in the range ($z,z+\Delta z$). 
The quantity $\Delta N_\text{approx}$ has the physical sense of the fluctuation mean and depends on quality of the LSM approximation as well as the approximation formula.

One can introduce an index $i$ denoting the ordinal number of the bin for the $\Delta N$ and $\delta$ values. Thus, $\delta_i$ is a relative difference between the observed number of galaxies and the theoretically expected one derived by the chosen approximation function of homogeneous distribution of galaxies in the $i$-bin.

We take the sequence of fluctuations $\delta_i$ only with either positive or negative values as a criterion for detecting structure in the fluctuation pattern. 
So, we can introduce a new index $j$ as the ordinal number of the detected structure. 
We denotes the middle of first bin of each such sequence as $z_\text{start}$ and the middle of last bin as $z_\text{final}$.
Each $j$-structure has the final bin, which is the start bin for $j+1$-structure except for the last one. We denotes all such transition bins as $j$-bins.
The last structure with the index $j=m+1$ has $z_\text{final} \approx z_\text{max}$, where $z_\text{max}$ is the size of galaxy sample. We exclude this structure from analysis. Thus, we detect $m$ structures that we now consider as a new random process of candidates for large-scale structures of galaxies.

We introduce quantity $ \delta_j $ as the average value over all fluctuations $ \delta_i $ of $j$-structure by the equation
\begin{equation}
    \delta_j = \sum \limits_{i \in j} \delta_i / n \, ,
\end{equation}
where $ n $ is the number of $ i $-bins inside $ j $-structure. 
In this paper, we consider structures with $n>2$. This leads to the fact that the redshift size of detectable structures is greater than $\Delta z=0.2$ in redshift radial space.
Further, we calculate an unbiased estimate of the variance of observed fluctuations of the number of galaxies $s_j^2$ and a variance of the sample mean $\hat{\sigma}_{\delta_j}^2$ for each structure by the equations~\citep[p.~670, Eqs.~19.2-8]{Korn1968}
\begin{equation}
    s_j^2 = {\sum \limits_{i \in j} (\delta_i-\delta_j)^2} / {(n-1)} \, ,
    \label{eq:s_obs}
\end{equation}
\begin{equation}
    \hat{\sigma}_{\delta_j}^2 = \frac{s_j^2}{n} \, ,
    \label{eq:sigma_obs}
\end{equation}
where $ n $ is the number of $ i $-bins inside $j$-structure.
From now on, we will use designations $\sigma$ instead of $\hat{\sigma}$ as estimate of variance and $\sigma_\text{obs}$ instead of $|\delta_j|$ as the mean value of observed fluctuations of $j$-structure. 

The mean value of observed fluctuations over all structures $\sigma_\text{mean}$ can be found as follows
\begin{equation}
    \sigma_\text{mean}=\overline{\sigma_\text{obs}} = \sum \limits_{j=1}^m \frac{ |\delta_j| } { m }  = \sum \limits_{j=1}^m \frac{\sigma_\text{obs} }{ m } \, ,
\end{equation}
where $ m $ is the number of detected $j$-structures.

We can obtain an error of the mean value of observed fluctuations over all structures $\sigma_{\sigma_\text{mean}}$, taking into account the Poisson noise, 
by the formula
\begin{equation}
    \sigma^2_{\sigma_\text{mean}} = \sigma^2_{ \overline{\sigma_\text{obs}} } + \frac{({ \overline{\sigma_\text{P}} })^2}{m}  \, ,
    \label{eq:sigma_mean}
\end{equation}
where 
$\sigma_{ \overline{\sigma_\text{obs}} }$ is the standard deviation of $\sigma_\text{mean}$, obtained similarly to $\sigma_{\sigma_\text{obs}}$ by Eqs.~(\ref{eq:s_obs}) and~(\ref{eq:sigma_obs}),
${ \overline{\sigma_\text{P}} }$ is the average Poisson noise over all structures,
and $ m $ is the number of detected $j$-structures
. 
Such approach to Poisson errors impairs the estimate of the mean error of the observed cosmic variance, but allows one to detect possible structures in a sample with a signal-to-noise ratio less than unity (see, for example, Figs.~6 and~7 from~\citet{Shirokov2016}). We can do this because the amplitude of the Poisson errors varies weakly within a single sample of galaxies with a factor of about 2--3. 

We rename $\sigma_\text{obs}$ as an amplitude of fluctuations of the observed number of galaxies or just the fluctuation amplitude in the context of one structure and $\sigma_\text{mean}$ as an average fluctuation amplitude over all detected structures in the context of the galaxy sample.

\subsection{Comparison with the SCM predictions}

According to the $\Lambda$CDM model, the theoretical cosmic variance of the fluctuations in the density of matter $\sigma^2$ for each redshift bin $(z, z+\Delta z)$ is equal to the sum of two variances,
\begin{equation}
    \sigma^2(z,\Delta z)=\sigma_\text{gal}^2+\sigma_\text{p}^2 \, ,
    \label{eq:sigma}
\end{equation}
where $\sigma_\text{p}^2$ is the classical Poisson noise, and $\sigma_\text{gal}^2$ is the variance
calculated by the formula
\begin{equation}
    \sigma_\text{gal}^2(V)=\frac{1}{(1+z)V^2}\int_V \text{d}V_1\int_V \text{d}V_2\xi_\text{gal}(|\mathbf{r}_1-\mathbf{r}_2|) \, ,
    \label{eq:sigma_gal}
\end{equation}
where $V=V(z,\Delta z)$ is the integration volume, $\xi(|\mathbf{r}_1-\mathbf{r}_2|)$ is the spatial two-point correlation function of matter density, and the factor $1/(1+z)$ takes into account the linear growth over time of the fluctuations of LSSU~\citep{Nabokov2010a}. In fact, this formula corresponds to the complete correlation function of matter (visible and dark). However, within the SCM frameworks, the one is derived only for dark matter, which is associated with the visible matter via the hypothesis of galaxy bias~\citep{Moster2011}. Various parameters of galaxies should be taken into account in order to better match the theory of dark matter with observations of baryonic matter~\citep{Moster2011}.

The variance of density fluctuations of dark matter is given by Eqs.~(10) and~(12) from~\citet{Moster2011} or by the formula
\begin{equation}
    \sigma_\text{dm}(z,\Delta z) = \frac{\sigma_a}{z^\beta+\sigma_b} \sqrt{\frac{0.2}{\Delta z}} \, ,
    \label{eq:sigma_dm}
\end{equation}
where parameters $ (a,\,b,$ and $\beta) $ are related to the angular dimensions of the COSMOS field and are equal to (0.069, 0.234, and 0.834), respectively~\citep{Moster2011}.

The galaxy bias has theoretical and observed values, which may differ from each other. For example, the accounting for the stellar mass of visible galaxies reduces the bias difference (as difference between theoretical and observed values)
with the factor of 2--3~\citep{Shirokov2016}. Both values are defined as a ratio of the correlation function of visible matter to the one of dark matter. The ratio of ``visible variance'' to ``dark variance'' can be used as a calculable approximation of galaxy bias, $b^2(z, \, m_\star, \, ...) = \xi_\text{gal} / \xi_\text{dm} \approx \sigma^2_\text{gal} / \sigma^2_\text{dm}$, which can be estimated from observations (with adding the ``obs'' prefix) and corrected for Poisson noise~\citep{Shirokov2016}. Galaxy bias is a complicated function that depends on a large number of parameters and redshift effects. If the observed function of galaxy bias is coincided with the theoretical one, $b_\text{obs}(z) / b(z, \, m_\star, \, ...) \approx 1$, then the bias hypothesis is confirmed. 

In this work, we calculate only the observed bias value $b_\text{obs}(z)$. The average bias $b_\text{mean}$ (as the output method's parameter) demonstrates the effect of difference between the observed variance of visible matter and theoretical variance of dark matter in the galaxy sample. We calculate the average bias value over structures as follows
\begin{equation}
    b_\text{mean} = \overline{b} = \frac{\sum b_j}{m} \, ,
\end{equation}
where $ m $ is the number of structures detected in the fluctuation pattern. We take an error for $ b_j $ as the relative error of the value $ \sigma_{obs} $ and its error, respectively.
The estimate of error $ \sigma_{b_\text{mean}} $ is the standard deviation of $ b_\text{mean} $. 

According to the formula~(\ref{eq:sigma}), it is necessary to take into account the Poisson noise ($\sigma^2_\text{P} = 1 / \Delta N$) when we calculate the observed bias value. The noise may be too high on poor samples (with $\Delta N<10^4$). 
A poor sample may give a small fluctuation amplitude $\delta_\text{obs}(z, \Delta z)$ relatively to a large Poisson background $\sigma_\text{P}$ that will lead to a negative observed variance $\sigma_\text{obs}$. 
In this paper, we eliminate such structures.  

\subsection{Alternative approach}

The observed distribution of galaxies is consistent with the power-law nature of the correlation function of matter density $\xi_\text{pl}(r)$ = $(r_0/r)^\gamma$ (e.g.,~\citet{Shirokov2016,Tekhanovich2016}). Using Eq.~(\ref{eq:sigma_gal}), one can calculate a power-law estimate of the cosmic variance ($\sigma_\text{gal} = \sigma_\text{pl}$) with the parameters $\gamma=1$, $r_0=5$ from~\citet{Shirokov2016}
as follows
\begin{equation}
    \sigma_{pl}^2(V,r_0,\gamma)=\frac{1}{(1+z)V^2}\int_V \text{d}V_1\int_V \text{d}V_2\xi(|\mathbf{r}_1-\mathbf{r}_2|,r_0,\gamma) \, ,
\end{equation}
and compare its with the observed variance corrected for Poisson noise by introducing the power-law bias function $ b_\text{pl}(z) = \sigma_\text{obs}/\sigma_\text{pl} $.
The consistency between the power-law and observed bias functions, $b_\text{obs}(z) / b_\text{pl}(z) \approx 1$, will shows the efficiency of applying the power-law correlation function (PLCF) without taking into account the influence of different selection effects for estimating the observed cosmic variance. This conclusion may be important for future testing of new cosmological models.
We also get the average value of the power-law bias function $\overline{b_\text{pl}}$ as the output method's parameter and its standard deviation $\sigma_{\overline{b_\text{pl}}}$.

\subsection{Calculating the fluctuation scales}

The metric distance in terms of redshift in the SCM is given by the formula
\begin{equation}
    r(z)_\text{Mpc} = \frac{c}{H_0} \int \limits_0^z \left(     \Omega_\text{v}+\Omega_\text{m}^3(1+z) \right)^{-\frac{1}{2}} \text{d}z \, ,
    \label{eq:size}
\end{equation}
where $c$ is the speed of light, $H_0 = 70$ km\,s$^{-1}$\,Mpc$^{-1}$ is the Hubble constant, the vacuum density (dark energy) $\Omega_\text{v} =  0.7$, the matter density (visible and dark) $\Omega_\text{m} = 1 - \Omega_\text{v} = 0.3$, and $z$ is the source redshift. Since we are working in the co-moving space we will use the metric distance, not the luminosity distance.

Here, we introduce a more stringent criterion to estimate the linear scales of large-scale structures than it was used in the papers by~\citet{Nabokov2010a,Shirokov2016}. 
As it is noted above, we calculate the structure borders $z_\text{start}$ and $z_\text{finish}$ as the middle of $j$-bins, where the fluctuation function $\delta_i$ changes a sign.
The structure size $R_j$ is given by the equation
\begin{equation}
    R_j = \Delta r(z,\Delta z) = r( z^\text{finish}_{j} ) - r ( z^\text{start}_{j-1} ), j\in(1,2, ... ) \, ,
    \label{eq:structure_size}
\end{equation}
where the values $z_j$ and $z_{j+1}$ correspond to the midpoints of the $j$-bins ($z^\text{start}_{0}=0$). We define the upper and lower errors of the structure size as half of the corresponding metric sizes of the start and final $j$-bins.

Since we are interested in average values of the target variables for a galaxy sample, we also calculate the average size of the detected structures $R_\text{mean} = \overline{R_j}$. The errors of $R_\text{mean}$  take into account both the root-mean-square deviation from the mean, and the half sizes of the boundary bins by the formula
\begin{equation}
    \sigma^2_{R_\text{mean}} = \sigma^2_{\overline{R_j}} + (\overline{\sigma_{R_j}})^2 \, ,
\end{equation}
where $\sigma_{ \overline{R_j} }$ is the standard deviation of $R_\text{mean}$, and $\overline{\sigma_{R_j}}$ is the mean of upper (or, respectively, lower) $R_j$ errors.

According to the SCM, the correlation function of spatial density of dark matter is equal to zero at the scale of $r_0=174$ Mpc, $\xi_\text{gal}(174\,\text{Mpc} )=0$, without reference to the galaxy bias value~\citep{SylosLabini2008}.
The bins with sizes exceeding this value can be considered as independent and requiring a sign change of the observed fluctuations $\delta_i=\delta_\text{obs}(z,\Delta z)$ on a scale about the double one $2r_0 \sim 350$ Mpc. 
Modern cosmological N-body simulations, such as the Horizon Run 2 simulation, predict inhomogeneities with a size of about 300 Mpc for the brightest galaxies, which are modelled over dark matter halos~\citep{Park2012}.
The BLUETIDES simulation, in which galaxies are modelled by simulating gas, shows
the clustering galaxies at high redshifts
with the galaxy bias $ b \approx 8 $~\citep{Bhowmick2018}.

A comparison of metric distances $r(z)$ (comoving space) and luminosity distances $d_\text{L}=(1+z)r(z)$ (proper space) is shown in Table~\ref{tab:bin_scales} for redshift bins $\Delta z = 0.1$, which are denoted by the midpoints $\overline{z}$. 
The difference between the corresponding distances 
caused by the factor $(1+z)$.

\begin{table*}\centering 
    \begin{tabular*}
    {\textwidth}
{@{\extracolsep{\fill}}cccccccccccc}
    \hline
        \hline
$\overline{z}$ & 0.05	&	0.95	&	1.45	&	1.55	&	1.65	&	1.95	&	2.95	&	3.95	&	4.95	&	5.95	\tabularnewline
        \hline
$\Delta r(z, \Delta z)$, Mpc & 407	&	244	&	184	&	{\bf 175}	&	166	&	144	&	95	&	68	&	52	&	41	\tabularnewline
$\Delta d_\text{L}(z, \Delta z)$, Mpc & 448	&	784	&	866	&	879	&	890	&	920	&	989	&	1032	&	1063	&	1087	\tabularnewline

        \hline
    \end{tabular*}                     
    \caption{The metric size of bins $\Delta z = 0.1$ in Mpc at various redshifts. $\Delta r(z)$ is the size in the moment $t=t(z)$ (comoving distances), and $\Delta d_\text{L}$ is the size in the moment $t=t(0)$ (proper distances).}
    \label{tab:bin_scales}
\end{table*}

\subsection{Applications of the method}

The methods of LSSU analysis should be maximally reliable and robust for any geometries. This way, for example, the sample depth should not significantly affects the identified structures, as we show in Appendix A for two disjoint samples of the COSMOS and UltraVISTA galaxies from the COSMOS2015 catalogue.
Besides, though we did not require correlation between two disjoint samples, one can see a correlation for several structures at small redshifts that indicates their presence.

It is possible to consider density fluctuations of matter in metric space (with and without taking into account the temporal effects) instead of redshift space. In such configuration, the uniform radial density distribution grows as a power-law function of the radius and, after reaching a maximum, also falls as one (with a negative exponent). However, for the transition from redshifts to metric distances, it is necessary to accept a certain cosmological model.
We get different distances and cosmic variance in different cosmological models.
It means that we need to calculate the fluctuation tables in a grid of models with a search for the optimum cosmological parameters. 
An illustration of a metric radial histogram in the $ \Lambda $CDM model frameworks with a uniform approximation for the UltraVISTA subsample is given in Appendix B.

\section{Results}
    \label{sec:Results}

\subsection{Entire sample of the COSMOS2015 catalogue}

\begin{figure*}  \centering

    \includegraphics[width=0.49\linewidth]{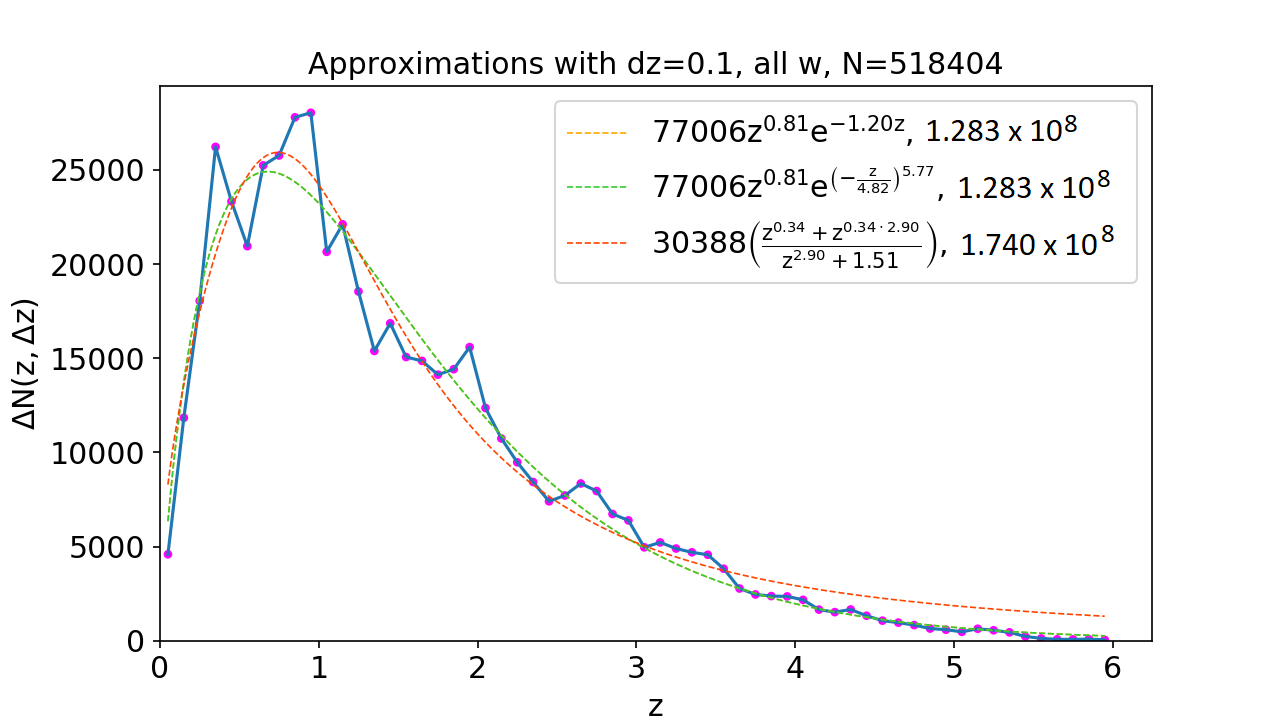}
    \hfill
    \includegraphics[width=0.49\linewidth]{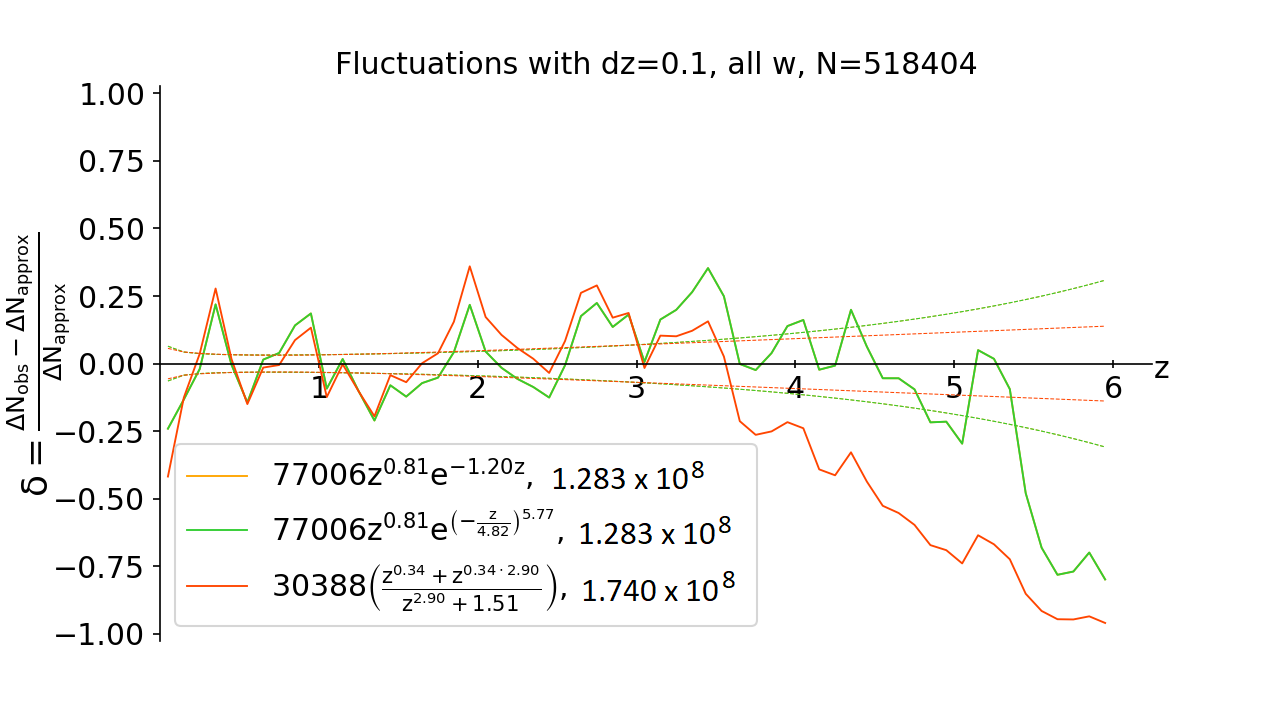}

    \caption{
    \emph{Left:} Histogram of the radial distribution of photometric redshifts from the COSMOS2015 catalogue in the range $ 0 < z < 6 $ for 518,404 galaxies within bins $\Delta z = 0.1$, and the approximations. Dotted lines mark the least squares fittings.
    \emph{Right:} The fluctuation pattern. Dotted lines mark Poisson noise 5$\sigma$.
    }
    \label{fig:COSMOS2015_w=all_dz=0.1}
    
\end{figure*}

Fig.~\ref{fig:COSMOS2015_w=all_dz=0.1} (left) shows the radial histogram of the number of galaxies in the COSMOS2015 catalogue (without additional sampling) at all redshifts $ 0 < z < 6 $ for 518,404 galaxies. The LSM approximations are shown by dashed lines: red corresponds to the theoretical formula~(\ref{eq:N3(z)}), green to the empirical formula (\ref{eq:N1(z)}), orange to its simplified form of Eq.~(\ref{eq:N2(z)}). 
The legend presents approximation formulas with coefficient values, and the root of the residual sum. The fluctuation pattern is shown in the right panel of Fig.~\ref{fig:COSMOS2015_w=all_dz=0.1}. 
Our LSM analysis gave a coincidence of the green and orange curves with high accuracy so that they can be considered as equivalent ones.

On the one hand, as it can be seen in the figure, the curves of the empirical cases coincide. This implies that the formula (\ref{eq:N2(z)}) is successfully simplified. However, the presence of two parameters in exponent leads to the uncertainty of determining the coefficients, which can result in the nonphysical large values of parameters. 
On the other hand, a power-law theoretical curve corresponding to the formula (\ref{eq:N3(z)}) has the similar behaviour at small redshifts, but describes the histogram at high redshifts poorly.

At small redshifts, $ z < 1.5  $, both approximations give a mutually consistent result with an average fluctuation amplitude of $ \delta \approx 10 $ \% and an average structure scale of $\Delta R \approx 700 $ Mpc. At intermediate redshifts, $ 1.5 < z < 3.5 $, a difference between the empirical and theoretical approximations is increasing. At large redshifts, $ 3.5 < z < 6 $, the difference grows rapidly, as well as a difference between the histogram and the approximations themselves that indicates an unsuccessful approximation of the histogram at these scales. This effect can be related with a lack of statistics at highest redshifts or with an unknown redshift systematics.

Nevertheless, in the fluctuation pattern, the regions of deficiency and excess of galaxies with a scale of 3--5 redshift bins are clearly visible. It can be noted that the fluctuations $\delta_i$ of the empirical curves are more stable with respect to redshift around $ \delta \sim 10 $ \% in the range $ 0 < z < 5 $. 
That emphasises an importance of choosing the distribution maximum and a successful approximation of the distribution tail.

The fluctuation pattern, obtained from the analysis of the entire COSMOS2015 catalogue, is similar to periodic density oscillations, which can be also seen in the deep-UltraVISTA sample in Appendix B (in metric space). 

\subsection{The \emph{w}-sampling}

To account for the photometric redshift uncertainties, we introduce the sampling parameter $ w $, which is equal to the difference of one and the relative error in determining $ z $,
\begin{equation}
    w_i = 1 - \frac{\sigma_i(z)}{z_i}, \, i=1,\,2,\,...\,,\,N\,,
    \label{eq:w}
\end{equation}
where $\sigma_i(z)$ is 1$\sigma$ uncertainty of each redshift $z_i$  and $N = 518,404$. Based on the choice of the parameter $ w $, we considered 4 samples from the COSMOS2015 catalogue:

\begin{itemize}
    \item any $ w $ (no sampling), i.e. the entire sample;
    \item $ w > 0.7 $ (weak sampling), i.e. galaxies with relative error $ \sigma (z) < 30 \% $ at significance level of $ 1 \sigma $ or $ \sigma (z) < 90 \% $ at significance level of $ 3 \sigma $;
    \item with $ w > 0.9 $ (medium sampling), i.e. galaxies with relative error $ \sigma (z) < 10 \% $ at significance level of $ 1 \sigma $ or $ \sigma (z) < 30 \% $ at significance level of $ 3 \sigma $;
    \item with $ w > 0.97 $ (strong sampling), i.e. galaxies with relative error $ \sigma (z) < 3 \% $ at significance level of $ 1 \sigma $ or $ \sigma (z) < 9 \% $ at significance level of $ 3 \sigma $.
\end{itemize}

\begin{figure*}    \centering

    \includegraphics[width=0.49\linewidth]{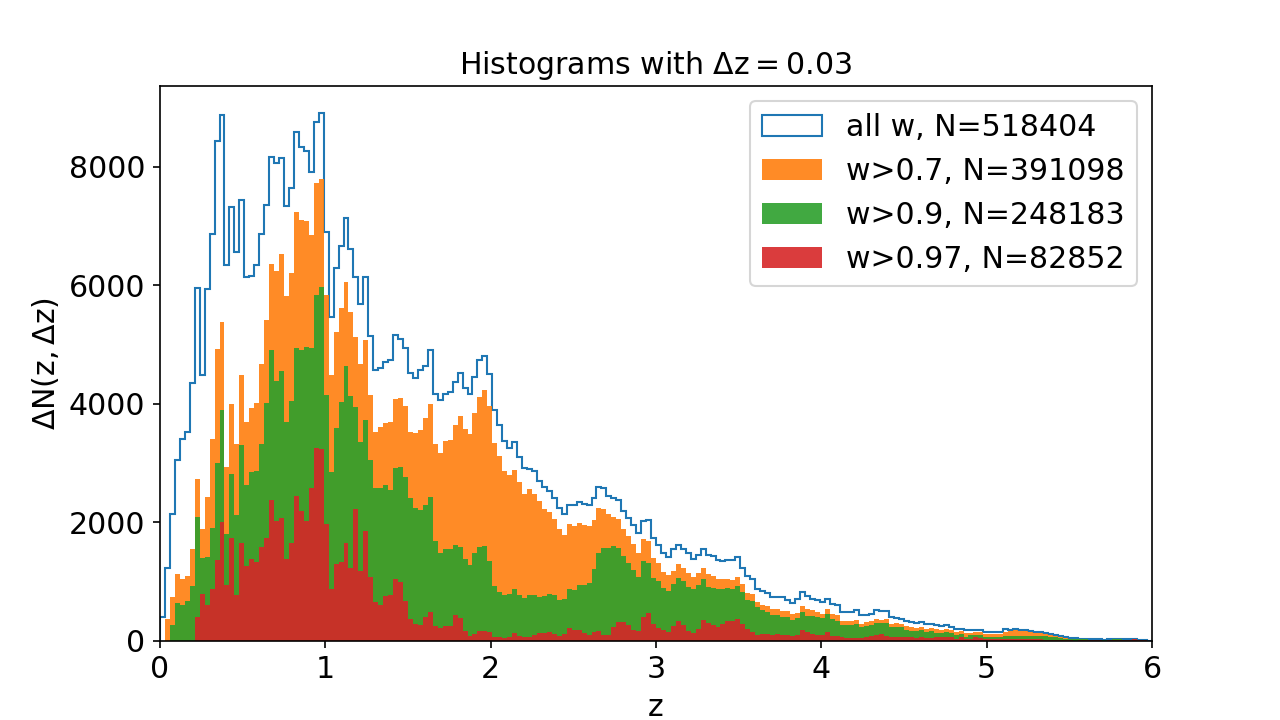}
    \hfill
    \includegraphics[width=0.49\linewidth]{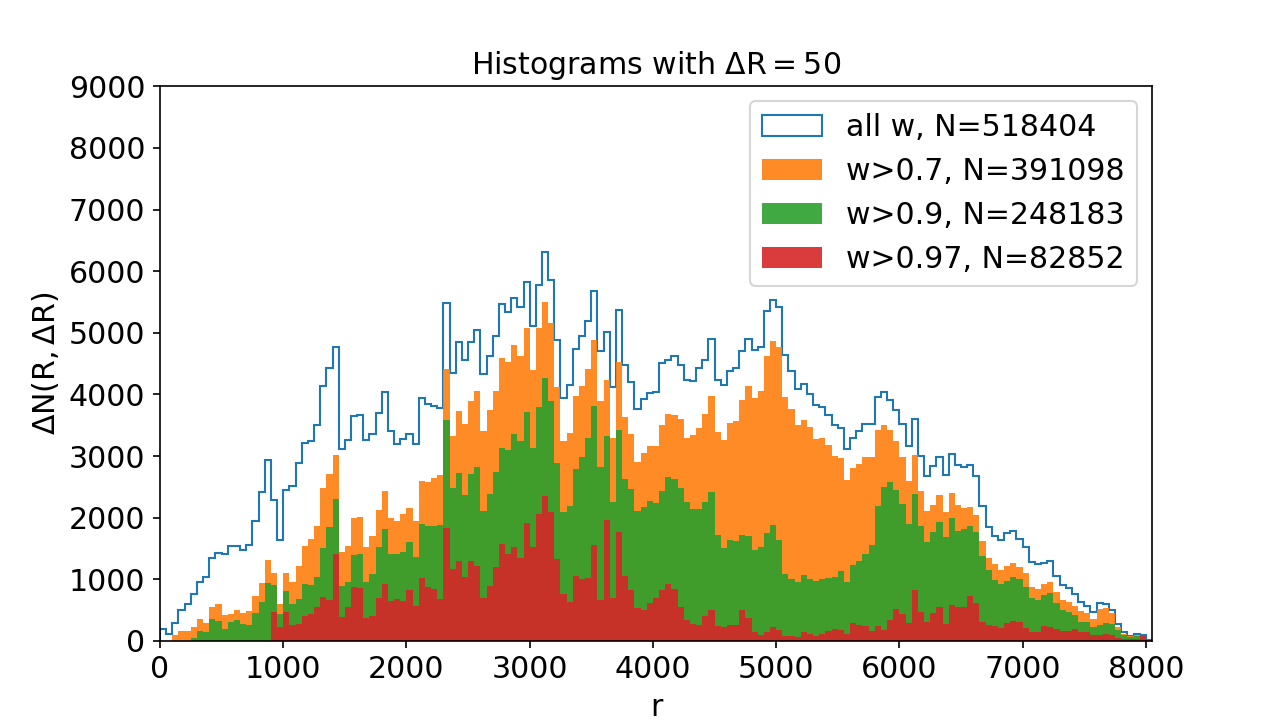}
    
    \caption{
    High resolution differential histograms of radial distribution of galaxies from the COSMOS2015 catalogue in redshift space (left) with $\Delta z=0.03$, and metric space (right) with $\Delta r(z) = 50$ Mpc for the various values of the sampling parameter $w$.}
    \label{fig:histogram_w}
\end{figure*}

The high resolution histograms of the radial distribution of the COSMOS2015 galaxies for the selection parameter $w$ are shown in Figure~\ref{fig:histogram_w}. 
The left panel demonstrates how the selection by the quality of photometric redshifts reduces the number of galaxies and reveals the large-scale structures, for example, at $ 2 < z < 4 $ in redshift space. 
The right panel is the left plot but in metric space, where distance units are in Mpc. We can see various structures (on Gpc scales) as a sum of the physics of the LSSU and observational selection effects.

\begin{figure*}  \centering
    
    \includegraphics[width=0.49\linewidth]{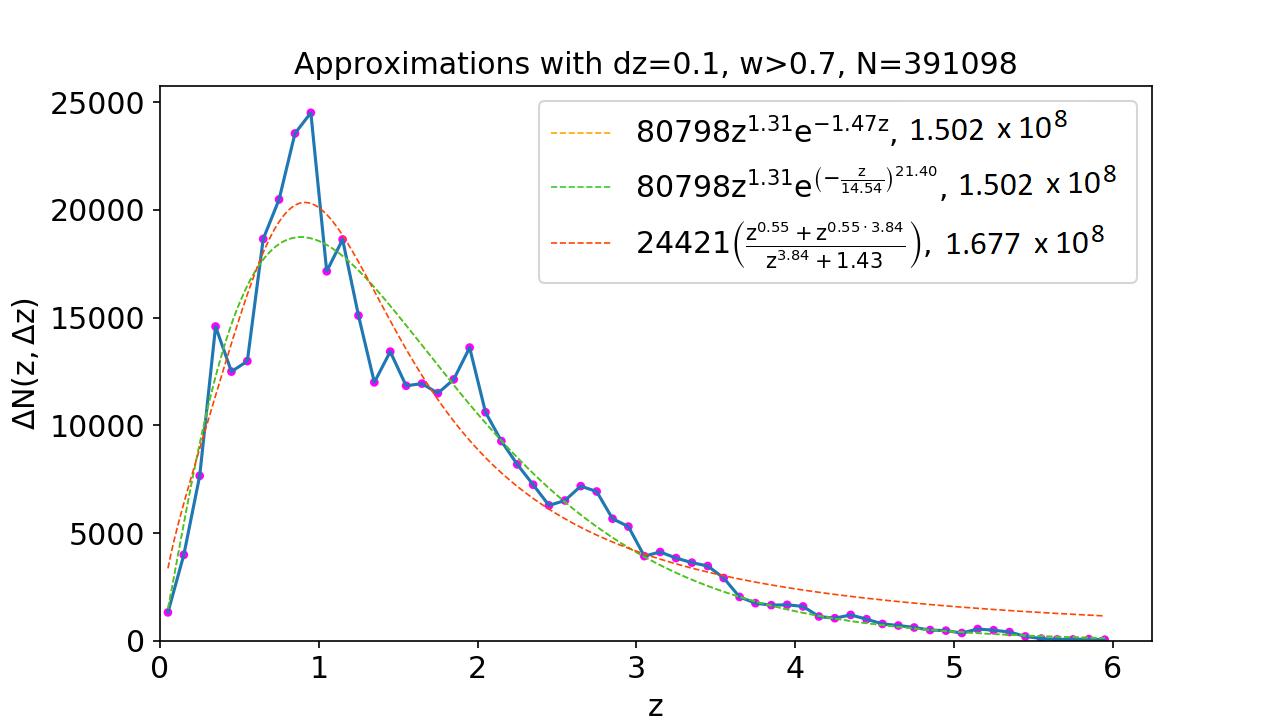}
    \hfill
    \includegraphics[width=0.49\linewidth]{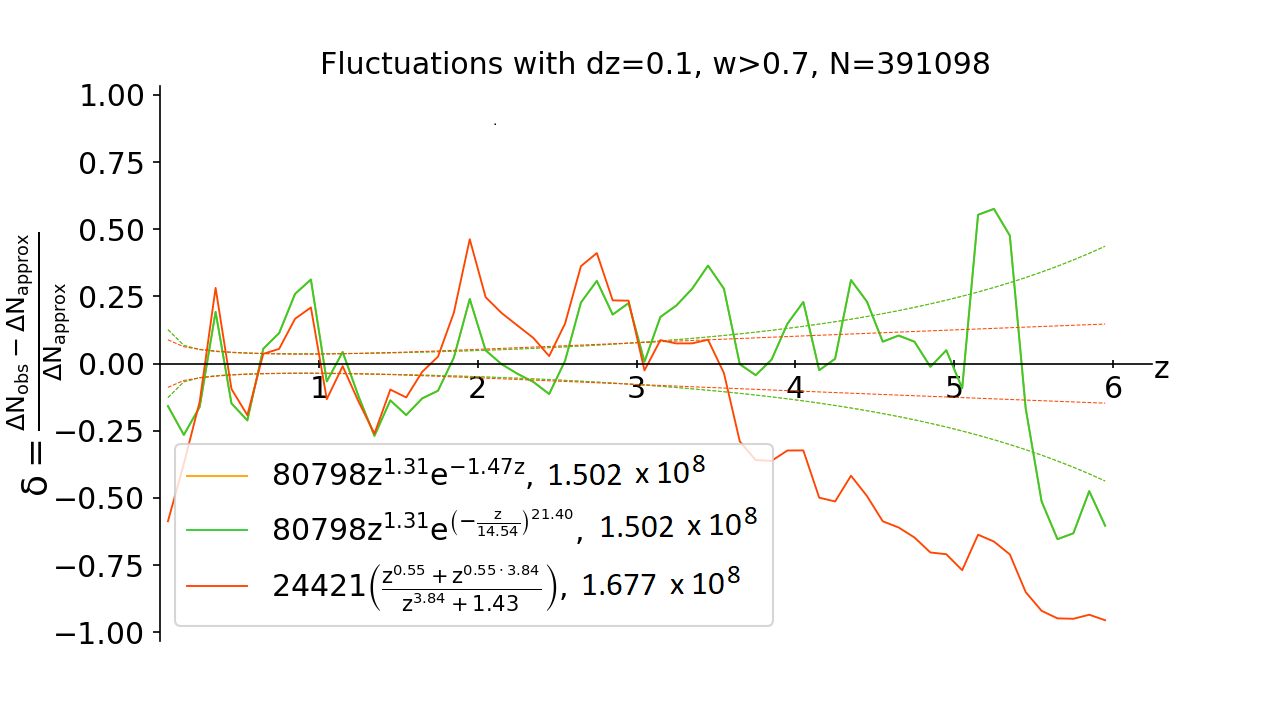}
    \vfill
    \includegraphics[width=0.49\linewidth]{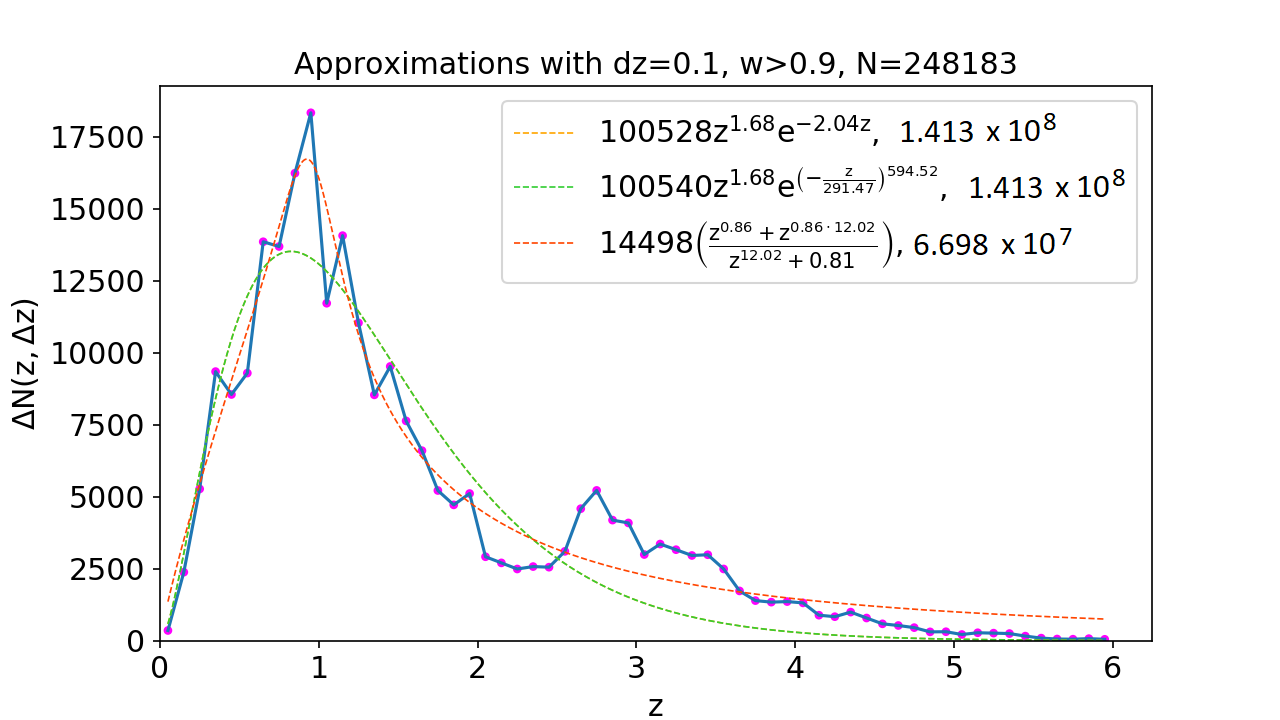}
    \hfill
    \includegraphics[width=0.49\linewidth]{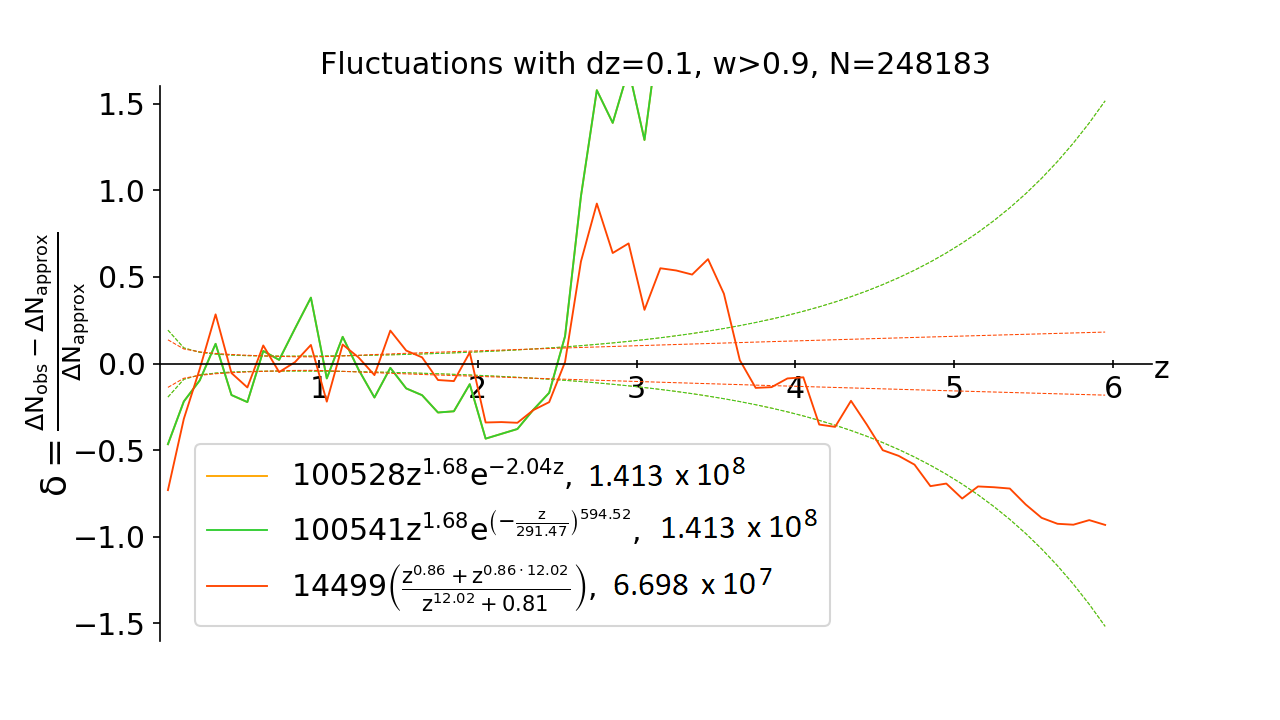}
    
    \caption{Histograms of radial distribution of the photometric redshifts and corresponding fluctuation patterns for samples with the parameter $ w = 0.7 $ for 391,098 galaxies (top), and $ w = 0.9 $ for 248,183 galaxies (bottom) with a bin $ \Delta z = 0.1 $. The dashed lines indicate the least squares fit (left) and Poisson noise 5$ \sigma $ (right).}
    \label{fig:COSMOS2015_w=0.7_and_0.9_dz=0.1}
    
\end{figure*}

Figure~\ref{fig:COSMOS2015_w=0.7_and_0.9_dz=0.1} (left panels) shows the histograms of the radial distribution of galaxies for sampling parameters $ w = 0.7 $ and $ w = 0.9 $ and demonstrates that reducing the sample based on the quality of photometric redshift helps to reveal the LSSU, for example, at $ 2 < z < 4 $ with scale of about Gpc. 
In this way, the method can be also applied not only in the redshift space, but also in the metric space (see Appendix B). 

For weak sampling ($w=0.7$), the fluctuation pattern differs slightly from the entire sample, and the number of galaxies is about 75\% of the total number. 
The empirical approximation describes the observed histogram better up to $ z \approx 5 $, while the theoretical curve is successful only up to $ z \approx 3 $. 

For strong sampling ($w=0.9$), the large-scale structures are seen more clearly, and the number of galaxies is about 48\% of the total number. 
The empirical approximation describes the observed histogram poorly, while the theoretical curve has the smallest sum of residual squares. 
In this case, the region of galaxy deficiency (void) in the redshift range $ 2 < z < 2.5 $, and the region of galaxy excess (supercluster) in the range $ 2.5 < z < 3.5 $ are clearly visible.

It is important to note that for samples with more strictly selected photometric redshifts, the empirical formula (\ref{eq:N2(z)}) predicts an underdensity at large redshifts, while the theoretical the formula~(\ref{eq:N3(z)}) gives the best result.

We present the structure parameters detected by the method from the corresponding fluctuation patterns for the four values of $ w $ ($ w $ = ``all'' means the case of no sampling) in Table~\ref{tab:COSMOS2015_w=all_dz=0.1_EMPIRY} for empirical approximations of homogeneity (that are coincided),
and in Table~\ref{tab:COSMOS2015_w=all_dz=0.1_THEORY} for the theoretical curve. The description of structure table contents in detail is given in Section~\ref{sec:The Methods of Space Structure Analysis} and Appendix A.

The structure tables indicate the existence of physical large-scale structures in Gpc scales. The largest structures found in the COSMOS2015 catalogue are detected at redshifts: $ 2 \lesssim z \lesssim 2.5 $ (void), and $ 2.5 \lesssim z \lesssim 3.5 $ (cluster).

The main output of the tables is the characteristic size of structures $ R_\text{mean} $ and their average characteristic amplitude $ \sigma_\text{mean} $. 
We obtain the galaxy bias functions $ b(z) $ and the average bias $ b_\text{mean} $ through comparison of the predicted cosmic variances in different models with the observations.  These parameters are convenient for fast comparison of different samples of galaxies.

\begin{table*}\centering \scriptsize
    \begin{tabular*}{\textwidth}
{@{\extracolsep{\fill}}ccccccccccccccccc} \hline \hline
        sample & $j$ & $n$ & $\overline{z}$ & $\Delta r(z)$, Mpc & $\sigma_\text{p}$ &  $\sigma_\text{obs}$ & $\sigma_\text{dm}$ & $b_\text{dm}$ & $\sigma_\text{pl}$ & $b_\text{pl}$ \\ \hline
        
        COSMOS2015 & 1 & 3 & 0.15 $\pm$ 0.10 & $774^{  +203}_{  -183}$ & 0.005 & 0.131 $\pm$ 0.064 & 0.188 & 0.7 $\pm$ 0.3 & 0.199 & 0.7 $\pm$ 0.3 \\ 
        $z_\text{max}=6$ & 2 & 3 & 0.35 $\pm$ 0.10 & $694^{  +183}_{  -163}$ & 0.004 & 0.065 $\pm$ 0.077 & 0.132 & 0.5 $\pm$ 0.6 & 0.179 & 0.4 $\pm$ 0.4 \\ 
        w=all & 3 & 3 & 0.55 $\pm$ 0.10 & $618^{  +163}_{  -145 }$ & 0.004 & 0.044 $\pm$ 0.050 & 0.104 & 0.4 $\pm$ 0.5 & 0.168 & 0.3 $\pm$ 0.3 \\ 
         & 4 & 5 & 0.85 $\pm$ 0.20 & $1035^{  +145}_{  -114}$ & 0.003 & 0.058 $\pm$ 0.049 & 0.097 & 0.6 $\pm$ 0.5 & 0.188 & 0.3 $\pm$ 0.3 \\ 
         & 5 & 8 & 1.50 $\pm$ 0.35 & $1265^{  +108}_{  -75}$ & 0.003 & 0.072 $\pm$ 0.028 & 0.073 & 1.0 $\pm$ 0.4 & 0.187 & 0.4 $\pm$ 0.2 \\ 
         & 6 & 4 & 2.00 $\pm$ 0.15 & $421^{  +75}_{  -65}$ & 0.004 & 0.072 $\pm$ 0.050 & 0.050 & 1.4 $\pm$ 1.0 & 0.151 & 0.5 $\pm$ 0.3 \\ 
         & 7 & 5 & 2.35 $\pm$ 0.20 & $482^{  +65}_{  -55}$ & 0.005 & 0.058 $\pm$ 0.022 & 0.048 & 1.2 $\pm$ 0.5 & 0.156 & 0.4 $\pm$ 0.1 \\ 
         & 8 & 12 & 3.10 $\pm$ 0.55 & $1000^{  +55}_{  -37}$ & 0.004 & 0.162 $\pm$ 0.033 & 0.045 & 3.6 $\pm$ 0.7 & 0.168 & 1.0 $\pm$ 0.2 \\ 
         & 9 & 5 & 3.95 $\pm$ 0.20 & $273^{  +36}_{  -32}$ & 0.010 & 0.059 $\pm$ 0.039 & 0.032 & 1.8 $\pm$ 1.2 & 0.140 & 0.4 $\pm$ 0.3 \\ 
         & 10 & 4 & 4.40 $\pm$ 0.15 & $180^{  +31}_{  -28}$ & 0.014 & 0.050 $\pm$ 0.055 & 0.028 & 1.7 $\pm$ 1.9 & 0.129 & 0.4 $\pm$ 0.4 \\ 
         & 11 & 7 & 4.85 $\pm$ 0.30 & $321^{  +28}_{  -24}$ & 0.013 & 0.126 $\pm$ 0.046 & 0.030 & 4.2 $\pm$ 1.5 & 0.145 & 0.9 $\pm$ 0.3 \\ \hline
        means: &  &  & 2.19 $\pm$ 0.22 & $642^{  +148}_{  -137}$ & 0.006 & 0.082 $\pm$ 0.012 & 0.075 & 1.6 $\pm$ 0.4 & 0.164 & 0.5 $\pm$ 0.1 \\ \hline \hline

        sample & $j$ & $n$ & $\overline{z}$ & $\Delta r(z)$, Mpc & $\sigma_\text{p}$ &  $\sigma_\text{obs}$ & $\sigma_\text{dm}$ & $b_\text{dm}$ & $\sigma_\text{pl}$ & $b_\text{pl}$ \\ \hline
        COSMOS2015 & 1 & 3 & 0.15 $\pm$ 0.10 & $774^{  +203}_{  -183}$ & 0.008 & 0.195 $\pm$ 0.035 & 0.188 & 1.0 $\pm$ 0.2 & 0.199 & 1.0 $\pm$ 0.2 \\ 
        $z_\text{max}=6$ & 2 & 3 & 0.35 $\pm$ 0.10 & $694^{  +183}_{  -163}$ & 0.005 & 0.039 $\pm$ 0.116 & 0.132 & 0.3 $\pm$ 0.9 & 0.179 & 0.2 $\pm$ 0.6 \\ 
         w=0.7 & 3 & 3 & 0.55 $\pm$ 0.10 & $618^{  +163}_{  -145}$ & 0.005 & 0.101 $\pm$ 0.080 & 0.104 & 1.0 $\pm$ 0.8 & 0.168 & 0.6 $\pm$ 0.5 \\ 
         & 4 & 5 & 0.85 $\pm$ 0.20 & $1035^{  +145}_{  -114}$ & 0.003 & 0.135 $\pm$ 0.069 & 0.097 & 1.4 $\pm$ 0.7 & 0.188 & 0.7 $\pm$ 0.4 \\ 
         & 5 & 8 & 1.50 $\pm$ 0.35 & $1265^{  +108}_{  -75}$ & 0.003 & 0.110 $\pm$ 0.036 & 0.073 & 1.5 $\pm$ 0.5 & 0.186 & 0.6 $\pm$ 0.2 \\ 
         & 6 & 4 & 2.00 $\pm$ 0.15 & $421^{  +75}_{  -65}$ & 0.005 & 0.078 $\pm$ 0.055 & 0.050 & 1.5 $\pm$ 1.1 & 0.151 & 0.5 $\pm$ 0.4 \\ 
         & 7 & 5 & 2.35 $\pm$ 0.20 & $482^{  +65}_{  -55}$ & 0.005 & 0.042 $\pm$ 0.022 & 0.048 & 0.9 $\pm$ 0.5 & 0.156 & 0.3 $\pm$ 0.1 \\ 
         & 8 & 12 & 3.10 $\pm$ 0.55 & $1000^{  +55}_{  -37}$ & 0.005 & 0.189 $\pm$ 0.035 & 0.045 & 4.2 $\pm$ 0.8 & 0.168 & 1.1 $\pm$ 0.2 \\ 
         & 9 & 3 & 3.75 $\pm$ 0.10 & $145^{  +37}_{  -35}$ & 0.013 & 0.010 $\pm$ 0.017 & 0.028 & 0.0 $\pm$ 0.0 & 0.118 & 0.0 $\pm$ 0.0 \\ 
         & 10 & 4 & 4.00 $\pm$ 0.15 & $202^{  +35}_{  -32}$ & 0.013 & 0.092 $\pm$ 0.059 & 0.030 & 3.0 $\pm$ 1.9 & 0.131 & 0.7 $\pm$ 0.4 \\ 
         & 11 & 7 & 4.55 $\pm$ 0.30 & $347^{  +31}_{  -26}$ & 0.014 & 0.116 $\pm$ 0.044 & 0.031 & 3.7 $\pm$ 1.4 & 0.147 & 0.8 $\pm$ 0.3 \\ 
         & 12 & 5 & 5.25 $\pm$ 0.20 & $193^{  +25}_{  -23}$ & 0.025 & 0.269 $\pm$ 0.164 & 0.026 & 10.3 $\pm$ 6.3 & 0.134 & 2.0 $\pm$ 1.2 \\ \hline
        means: &  &  & 2.37 $\pm$ 0.21 & $598^{  +144}_{  -134}$ & 0.009 & 0.115 $\pm$ 0.022 & 0.071 & 2.6 $\pm$ 0.8 & 0.160 & 0.8 $\pm$ 0.2 \\ \hline \hline

        sample & $j$ & $n$ & $\overline{z}$ & $\Delta r(z)$, Mpc & $\sigma_\text{p}$ &  $\sigma_\text{obs}$ & $\sigma_\text{dm}$ & $b_\text{dm}$ & $\sigma_\text{pl}$ & $b_\text{pl}$ \\ \hline

        COSMOS2015 & 1 & 3 & 0.15 $\pm$ 0.10 & $774^{  +203}_{  -183}$ & 0.010 & 0.260 $\pm$ 0.108 & 0.188 & 1.4 $\pm$ 0.6 & 0.199 & 1.3 $\pm$ 0.5 \\ 
        $z_\text{max}=6$ & 2 & 4 & 0.50 $\pm$ 0.15 & $955^{  +173}_{  -145}$ & 0.005 & 0.054 $\pm$ 0.086 & 0.124 & 0.4 $\pm$ 0.7 & 0.192 & 0.3 $\pm$ 0.5 \\ 
        w=0.9 & 3 & 5 & 0.85 $\pm$ 0.20 & $1035^{  +145}_{  -114}$ & 0.004 & 0.118 $\pm$ 0.080 & 0.097 & 1.2 $\pm$ 0.8 & 0.188 & 0.6 $\pm$ 0.4 \\ 
         & 4 & 3 & 1.15 $\pm$ 0.10 & $434^{  +114}_{  -102}$ & 0.005 & 0.011 $\pm$ 0.073 & 0.066 & 0.2 $\pm$ 1.0 & 0.149 & 0.1 $\pm$ 0.4 \\ 
         & 5 & 14 & 1.90 $\pm$ 0.65 & $1958^{  +102}_{  -55}$ & 0.003 & 0.196 $\pm$ 0.043 & 0.066 & 3.0 $\pm$ 0.7 & 0.189 & 1.0 $\pm$ 0.2 \\ \hline
        means: &  &  & 0.91 $\pm$ 0.24 & $1031^{  +302}_{  -287}$ & 0.005 & 0.128 $\pm$ 0.046 & 0.108 & 1.2 $\pm$ 0.5 & 0.183 & 0.7 $\pm$ 0.2 \\ \hline \hline
        
        sample & $j$ & $n$ & $\overline{z}$ & $\Delta r(z)$, Mpc & $\sigma_\text{p}$ &  $\sigma_\text{obs}$ & $\sigma_\text{dm}$ & $b_\text{dm}$ & $\sigma_\text{pl}$ & $b_\text{pl}$ \\ \hline
        COSMOS2015 & 1 & 3 & 0.35 $\pm$ 0.10 & $694^{  +183}_{  -163}$ & 0.011 & 0.290 $\pm$ 0.173 & 0.132 & 2.2 $\pm$ 1.3 & 0.179 & 1.6 $\pm$ 1.0 \\ 
        $z_\text{max}=6$ & 2 & 5 & 0.65 $\pm$ 0.20 & $1166^{  +163}_{  -129}$ & 0.006 & 0.089 $\pm$ 0.055 & 0.114 & 0.8 $\pm$ 0.5 & 0.196 & 0.5 $\pm$ 0.3 \\ 
        w=0.97 & 3 & 3 & 0.95 $\pm$ 0.10 & $487^{  +129}_{  -114}$ & 0.007 & 0.120 $\pm$ 0.229 & 0.075 & 1.6 $\pm$ 3.1 & 0.153 & 0.8 $\pm$ 1.5 \\ 
        & 4 & 3 & 1.35 $\pm$ 0.10 & $388^{  +102}_{  92}$ & 0.010 & 0.041 $\pm$ 0.135 & 0.059 & 0.7 $\pm$ 2.2 & 0.144 & 0.3 $\pm$ 0.9 \\ 
        & 5 & 9 & 1.85 $\pm$ 0.40 & $1216^{  +92}_{  -62}$ & 0.010 & 0.231 $\pm$ 0.074 & 0.064 & 3.6 $\pm$ 1.2 & 0.182 & 1.3 $\pm$ 0.4 \\ \hline
        means: &  &  & 1.03 $\pm$ 0.18 & $790^{  +227}_{  -212}$ & 0.009 & 0.154 $\pm$ 0.046 & 0.089 & 1.8 $\pm$ 0.5 & 0.171 & 0.9 $\pm$ 0.3 \\ \hline
    \end{tabular*}                     
    \caption{
    Tables of structures for $w$-samples from the COSMOS2015 catalogue for a bin $\Delta z = 0.1$ with $z_\text{max}=6$ and the various $\sigma_z$ selection. The approximation is by empirical formula (\ref{eq:N2(z)}).
    In the last string there are the means of corresponding values (by all structures): 
    mean of redshifts $z_\text{mean}$, 
    mean of structure sizes in Mpc $R_\text{mean}$, 
    mean of Poisson noise level $1\overline{\sigma_\text{P}}$,
    mean of observed cosmic variance $\sigma_\text{mean}$,
    mean of dark matter variance $\overline{\sigma_\text{dm}}$, 
    mean of dark matter bias $\overline{b_\text{dm}}$,
    mean of Peebles correlation function variance $\overline{\sigma_\text{pl}}$
    and its bias $\overline{b_\text{pl}}$ (see details in the text).
    }
    \label{tab:COSMOS2015_w=all_dz=0.1_EMPIRY}
\end{table*}

\begin{table*}\centering \scriptsize
    \begin{tabular*}{\textwidth}
{@{\extracolsep{\fill}}ccccccccccc} \hline \hline
sample & $j$ & $n$ & $\overline{z}$ & $\Delta r(z)$, Mpc & $\sigma_\text{p}$ &  $\sigma_\text{obs}$ & $\sigma_\text{dm}$ & $b_\text{dm}$ & $\sigma_\text{pl}$ & $b_\text{pl}$ \\ \hline
        COSMOS2015 & 1 & 3 & 0.15 $\pm$ 0.10 & $774^{  +203}_{  -183}$ & 0.005 & 0.169 $\pm$ 0.133 & 0.188 & 0.9 $\pm$ 0.7 & 0.199 & 0.9 $\pm$ 0.7   \\ 
        $z_\text{max}=6$ & 2 & 3 & 0.35 $\pm$ 0.10 & $694^{  +183}_{  -163}$ & 0.004 & 0.110 $\pm$ 0.084 & 0.132 & 0.8 $\pm$ 0.6 & 0.179 & 0.6 $\pm$ 0.5 \\ 
        w=all & 3 & 4 & 0.60 $\pm$ 0.15 & $900^{  +163}_{  -137}$ & 0.003 & 0.038 $\pm$ 0.038 & 0.112 & 0.3 $\pm$ 0.3 & 0.186 & 0.2 $\pm$ 0.2 \\ 
         & 4 & 4 & 0.90 $\pm$ 0.15 & $753^{  +137}_{  -114}$ & 0.003 & 0.023 $\pm$ 0.057 & 0.087 & 0.3 $\pm$ 0.6 & 0.174 & 0.1 $\pm$ 0.3 \\ 
         & 5 & 7 & 1.35 $\pm$ 0.30 & $1172^{  +114}_{  -83}$ & 0.003 & 0.076 $\pm$ 0.027 & 0.077 & 1.0 $\pm$ 0.4 & 0.186 & 0.4 $\pm$ 0.1 \\ 
         & 6 & 8 & 2.00 $\pm$ 0.35 & $989^{  +83}_{  -60}$ & 0.003 & 0.114 $\pm$ 0.041 & 0.059 & 1.9 $\pm$ 0.7 & 0.176 & 0.6 $\pm$ 0.2 \\ 
         & 7 & 7 & 2.75 $\pm$ 0.30 & $616^{  +57}_{  -45}$ & 0.005 & 0.134 $\pm$ 0.048 & 0.046 & 2.9 $\pm$ 1.1 & 0.161 & 0.8 $\pm$ 0.3 \\ 
         & 8 & 6 & 3.30 $\pm$ 0.25 & $420^{  +45}_{  -38}$ & 0.006 & 0.082 $\pm$ 0.026 & 0.039 & 2.1 $\pm$ 0.7 & 0.151 & 0.5 $\pm$ 0.2 \\ \hline
        means: &  &  & 1.42 $\pm$ 0.21 & $790^{  +155}_{  -137}$ & 0.004 & 0.093 $\pm$ 0.017 & 0.092 & 1.3 $\pm$ 0.3 & 0.177 & 0.5 $\pm$ 0.1 \\ \hline \hline
sample & $j$ & $n$ & $\overline{z}$ & $\Delta r(z)$, Mpc & $\sigma_\text{p}$ &  $\sigma_\text{obs}$ & $\sigma_\text{dm}$ & $b_\text{dm}$ & $\sigma_\text{pl}$ & $b_\text{pl}$ \\ \hline
        COSMOS2015 & 1 & 3 & 0.15 $\pm$ 0.10 & $774^{  +203}_{  -183}$ & 0.007 & 0.367 $\pm$ 0.129 & 0.188 & 2.0 $\pm$ 0.7 & 0.199 & 1.8 $\pm$ 0.7 \\ 
        $z_\text{max}=6$ & 2 & 3 & 0.35 $\pm$ 0.10 & $694^{  +183}_{  -163}$ & 0.005 & 0.015 $\pm$ 0.134 & 0.132 & 0.1 $\pm$ 1.0 & 0.179 & 0.1 $\pm$ 0.7 \\ 
        w=0.7 & 3 & 3 & 0.55 $\pm$ 0.10 & $618^{  +163}_{  -145}$ & 0.005 & 0.083 $\pm$ 0.066 & 0.104 & 0.8 $\pm$ 0.6 & 0.168 & 0.5 $\pm$ 0.4 \\ 
         & 4 & 5 & 0.85 $\pm$ 0.20 & $1035^{  +145}_{  -114}$ & 0.003 & 0.067 $\pm$ 0.060 & 0.097 & 0.7 $\pm$ 0.6 & 0.188 & 0.4 $\pm$ 0.3 \\ 
         & 5 & 8 & 1.40 $\pm$ 0.35 & $1335^{  +114}_{  -79}$ & 0.003 & 0.097 $\pm$ 0.032 & 0.076 & 1.3 $\pm$ 0.4 & 0.189 & 0.5 $\pm$ 0.2 \\ 
         & 6 & 14 & 2.40 $\pm$ 0.65 & $1561^{  +79}_{  -45}$ & 0.003 & 0.196 $\pm$ 0.039 & 0.055 & 3.5 $\pm$ 0.7 & 0.180 & 1.1 $\pm$ 0.2 \\ 
         & 7 & 6 & 3.30 $\pm$ 0.25 & $420^{  +45}_{  -38}$ & 0.007 & 0.044 $\pm$ 0.024 & 0.039 & 1.1 $\pm$ 0.6 & 0.151 & 0.3 $\pm$ 0.2 \\ \hline
        means: &  &  & 1.29 $\pm$ 0.25 & $920^{  +211}_{  -195}$ & 0.005 & 0.124 $\pm$ 0.046 & 0.099 & 1.4 $\pm$ 0.4 & 0.179 & 0.7 $\pm$ 0.2 \\ \hline \hline
sample & $j$ & $n$ & $\overline{z}$ & $\Delta r(z)$, Mpc & $\sigma_\text{p}$ &  $\sigma_\text{obs}$ & $\sigma_\text{dm}$ & $b_\text{dm}$ & $\sigma_\text{pl}$ & $b_\text{pl}$ \\ \hline
        COSMOS2015 & 1 & 3 & 0.15 $\pm$ 0.10 & $774^{  +203}_{  -183}$ & 0.010 & 0.360 $\pm$ 0.203 & 0.188 & 1.9 $\pm$ 1.1 & 0.199 & 1.8 $\pm$ 1.0 \\ 
        $z_\text{max}=6$ & 2 & 3 & 0.35 $\pm$ 0.10 & $694^{  +183}_{  -163}$ & 0.007 & 0.066 $\pm$ 0.109 & 0.132 & 0.5 $\pm$ 0.8 & 0.179 & 0.4 $\pm$ 0.6 \\ 
        w=0.9 & 3 & 3 & 0.55 $\pm$ 0.10 & $618^{  +163}_{  -145}$ & 0.006 & 0.029 $\pm$ 0.071 & 0.104 & 0.3 $\pm$ 0.7 & 0.168 & 0.2 $\pm$ 0.4 \\ 
         & 4 & 3 & 1.05 $\pm$ 0.10 & $459^{  +121}_{  -108}$ & 0.005 & 0.001 $\pm$ 0.109 & 0.070 & 0.0 $\pm$ 0.0 & 0.151 & 0.0 $\pm$ 0.0 \\ 
         & 5 & 4 & 1.50 $\pm$ 0.15 & $538^{  +97}_{  -83}$ & 0.006 & 0.059 $\pm$ 0.053 & 0.062 & 1.0 $\pm$ 0.9 & 0.160 & 0.4 $\pm$ 0.3 \\ 
         & 6 & 4 & 1.80 $\pm$ 0.15 & $463^{  +83}_{  -71}$ & 0.007 & 0.024 $\pm$ 0.043 & 0.054 & 0.4 $\pm$ 0.8 & 0.154 & 0.2 $\pm$ 0.3 \\ 
         & 7 & 7 & 2.25 $\pm$ 0.30 & $756^{  +71}_{  -55}$ & 0.006 & 0.205 $\pm$ 0.065 & 0.053 & 3.9 $\pm$ 1.2 & 0.168 & 1.2 $\pm$ 0.4 \\ 
         & 8 & 12 & 3.10 $\pm$ 0.55 & $1000^{  +55}_{  -374}$ & 0.006 & 0.483 $\pm$ 0.076 & 0.045 & 10.8 $\pm$ 1.7 & 0.168 & 2.9 $\pm$ 0.5 \\ \hline
        means: &  &  & 1.34 $\pm$ 0.19 & $663^{  +146}_{  -130}$ & 0.006 & 0.153 $\pm$ 0.064 & 0.089 & 2.7 $\pm$ 1.3 & 0.168 & 1.0 $\pm$ 0.4 \\ \hline \hline
sample & $j$ & $n$ & $\overline{z}$ & $\Delta r(z)$, Mpc & $\sigma_\text{p}$ &  $\sigma_\text{obs}$ & $\sigma_\text{dm}$ & $b_\text{dm}$ & $\sigma_\text{pl}$ & $b_\text{pl}$ \\ \hline
        COSMOS2015 & 1 & 3 & 0.55 $\pm$ 0.10 & $618^{  +163}_{  -145}$ & 0.008 & 0.005 $\pm$ 0.066 & 0.104 & 0.0 $\pm$ 0.0 & 0.168 & 0.0 $\pm$ 0.0 \\ 
        $z_\text{max}=6$ & 2 & 3 & 0.75 $\pm$ 0.10 & $548^{  +145}_{  -129}$ & 0.007 & 0.057 $\pm$ 0.067 & 0.087 & 0.7 $\pm$ 0.8 & 0.160 & 0.4 $\pm$ 0.4 \\ 
        w=0.97 & 3 & 4 & 1.20 $\pm$ 0.15 & $633^{  +114}_{  -97}$ & 0.008 & 0.063 $\pm$ 0.139 & 0.072 & 0.9 $\pm$ 1.9 & 0.166 & 0.4 $\pm$ 0.8 \\ 
         & 4 & 13 & 2.05 $\pm$ 0.60 & $1678^{  +92}_{  -53}$ & 0.008 & 0.338 $\pm$ 0.080 & 0.062 & 5.5 $\pm$ 1.3 & 0.185 & 1.8 $\pm$ 0.4 \\ 
         & 5 & 25 & 3.85 $\pm$ 1.20 & $1754^{  +53}_{  -25}$ & 0.014 & 1.104 $\pm$ 0.175 & 0.040 & 27.3 $\pm$ 4.3 & 0.170 & 6.5 $\pm$ 1.0 \\ 
         & 6 & 4 & 5.20 $\pm$ 0.15 & $147^{  +25}_{  -23}$ & 0.054 & 0.074 $\pm$ 0.087 & 0.025 & 2.1 $\pm$ 2.4 & 0.126 & 0.4 $\pm$ 0.5 \\ \hline
        means: &  &  & 2.27 $\pm$ 0.38 & $896^{  +290}_{  -282}$ & 0.017 & 0.273 $\pm$ 0.173 & 0.065 & 7.3 $\pm$ 4.4 & 0.162 & 1.9 $\pm$ 1.0 \\ \hline

    \end{tabular*}                     
    \caption{
    Tables of structures for $w$-samples from the COSMOS2015 catalogue for a bin $\Delta z = 0.1$ with $z_\text{max}=6$ and the various $\sigma_z$ selection. The approximation is by theoretical formula (\ref{eq:N3(z)}).
    In the last string there are the means of corresponding values (by all structures): 
    mean of redshifts $z_\text{mean}$, 
    mean of structure sizes in Mpc $R_\text{mean}$, 
    mean of Poisson noise level $1\overline{\sigma_\text{P}}$,
    mean of observed cosmic variance $\sigma_\text{mean}$,
    mean of dark matter variance $\overline{\sigma_\text{dm}}$, 
    mean of dark matter bias $\overline{b_\text{dm}}$,
    mean of Peebles correlation function variance $\overline{\sigma_\text{pl}}$
    and its bias $\overline{b_\text{pl}}$ (see details in the text).
    }
    \label{tab:COSMOS2015_w=all_dz=0.1_THEORY}
\end{table*}

The structure tables show that the observed cosmic variance can be described by the PLCF of density $\xi(r)$ = $(r_0/r)^\gamma$ with parameters $ \gamma = 1.0 $, $ r_0 = 5 $ at all redshifts~\citep{Nabokov2010a, Tekhanovich2016, Shirokov2016}. The PLCF bias $b_\text{pl}$ is greater than the dark matter bias $b_\text{dm}$ in factor about 2--3.

\section{Discussion}

As we note above, the galaxy bias is a complicated function that can rise with increasing redshift. A theoretical estimate of the dark matter galaxy bias function $b_\text{dm}(z)$ for the COSMOS field calculated by Eq.~\ref{eq:sigma_dm} can be improved by taking into account the stellar mass of the COSMOS galaxies~\citep{Shirokov2016}. Thereby our results are consistent with the SCM predictions. Nevertheless, the presence of fluctuations with a high value of the bias (5, 10, or 20) is a direct indication of the existence of large-scale structures that are interesting for the LSSU study. 

As we show on Figures~\ref{fig:COSMOS2015_w=0.7_and_0.9_dz=0.1} (bottom) and~\ref{fig:COSMOS_UltraVISTA_w=all_dz=0.1_zmax=6} (bottom), the larger value of the sampling parameter $ w $ leads to the increasing correlation of fluctuations between the COSMOS2015 and the UltraVISTA+deep-UltraVISTA samples. 
This means that an introducing additional sampling criteria for the quality of photometric redshifts of galaxies provides a more reliable fluctuation pattern, which was one of the goals of this work.

Comparison of the physical properties of the objects in the local Universe and at large redshifts requires new available observational spectral bands and instruments for multimessenger ranges (such as neutrino and gravitational waves). 
Observational cosmology based on multimessenger data allows one to verify existing cosmological models as well as formulating new ones~\citet{shirokov2020theseus}.
Our method can be used for analysis of photometric catalogs towards the transient objects (such as supernovae and gamma-ray bursts) detected by neutrino and gravitational-wave detectors.

The works by~\citet{Lovyagin2009, Nabokov2010b, Shirokov2017, Park2017} emphasise the importance of the analysis of the LSSU for the development of modern cosmology, which has become an increasingly important task in the 21st century.
The spatial distribution of galaxies reflects both the initial conditions in the early Universe and the evolution of primordial density perturbations. The analysis of fluctuations in the radial distribution of galaxies allows one to estimate the sizes and amplitudes of the largest structures in a given sample of galaxies.

Super-large fractal-like structures with scales of more than 100 Mpc reveal themselves both in the spatial distribution of galaxies in the local Universe at redshifts of $ z \lesssim 0.1 $ and in the quasars and gamma-ray bursts distributions at redshifts of $ z \sim 2 $~\citep{Gabrielli2005, Baryshev2012, Courtois2013, Einasto2016, Lietzen2016, Tekhanovich2016, Shirokov2017}.  
The analysis of fractal properties~\citep{Mandelbrot1982, Gabrielli2005, Baryshev2012} can be used to describe the properties of the large-scale distribution of matter~\citep{Tekhanovich2016, Shirokov2017}.

Note that in recent works there were detected several structures in the COSMOS field,
which also point to existence of large filamentary structures at $z\sim0.73$, called the COSMOS wall~\citep{Iovino2016}, structures at redshifts of $ 0.1 < z < 1.2 $~\citep{Darvish2017}, voids at $z\sim2.3$ ~\citep{Krolewski2018} and massive proto-supercluster at $z\sim2.45$~\citep{Cucciati2018}. The very large structure of the dark matter with size about 1,000 Mpc was detected in the COSMOS field by using method of the weak gravitational lensing 
\cite{Massey2007b, Massey2007}.

Our results demonstrated in this paper are consistent with our previous works and have a higher quality of detecting fluctuation structure in redshift space. The positive correlation of fluctuations between independent spectral and photometric surveys of different groups is shown in the work~\citet{Shirokov2016}.

\section{Conclusion}
    \label{sec:Conclusion}

We have performed the robust statistical analysis of new photometric catalogue COSMOS2015 and obtained new results, which confirm our preceding works.
We have considered the radial fluctuations of the number of galaxies along the line of sight in the photometric redshift space for the optical part of the COSMOS2015 catalogue and some of its subsamples. We have calculated the histograms of the number of galaxies in redshift bins, constructed empirical and theoretical approximations for the homogeneous distribution by using the LSM, and performed the comparative analysis of their quality. We have obtained the tables of structures for each sample and for each approximation. The bin size, $ \Delta z = 0.1 $, was chosen such as to maximize the spatial resolution (at a relatively low level of Poisson noise) and minimize the grid effect for determining photometric redshifts with errors $ \sigma_z < 0.021 (1 + z) $ at high redshifts $3<z<6$~\citep{Laigle2016}. 

Essential improvements in photometric redshift techniques, using the best SED fitting in the COSMOS field imaged in a large number of filters~\citep{Laigle2016} or deep learning methods~\citep{Pasquet2019}, allow to reach a redshift uncertainty
$\sigma_z = 0.007 (1+z)$ at small redshifts, which corresponds to a distance uncertainty of $\sim$ 40 Mpc (at $z\sim1$). 
So, in our paper, we present the firmly observed structures at redshift $z\sim 2$ with sizes $l \sim 700$ Mpc, which are larger than the BAO scale 
$l_{bao} \sim 100$ Mpc.

We have developed a Python software for the line-of-sight analysis based on works by~\citet{Lovyagin2009, Nabokov2010a, Nabokov2010b, Shirokov2016}. 
Our method takes into account the integral values within each bin for all quantities that increases the mathematical rigor and certainty of the results. 
In this paper, we calculate the structure means as the target variables of method and use algorithms for determining errors. These features make the method more robust for comparative analysis of different samples of galaxies.

Our analysis confirm the presence of dark matter structures from the paper by~\citet{Massey2007b} at small redshifts. Thus, the observations of visible matter and the observations of dark matter are consistent. Moreover, we have obtained the huge structures at high redshifts. This results are consistent with the $\Lambda$CDM model, if the corresponding bias is taken into account.

Based on the results of the work, we can draw the following conclusions:

\begin{itemize}

\item The method for analyzing radial fluctuations of the number of galaxies along the line of sight (see the last work~\citet{Shirokov2016}) now takes into account the integral values within each bin for all quantities. That increases the mathematical rigor and certainty of the results. The target variables  became more robust for comparative analysis of different samples of galaxies with the developed algorithms for errors estimation.
The use of logarithmic redshift bins can better take into account the photometric errors. Moreover, the metric bin size in logarithmic scale slightly depends on the redshift. 
Instead of analytical bias functions, numerical estimates of galaxy biases can be obtained from N-body simulations of Universe (in a grid of model parameters) or by using the concept of model fractal catalogs, as in~\citet{Shirokov2017}.

\item For the case of the theoretical form of approximation of homogeneity in the $\Lambda$CDM frameworks, the average standard deviation of detected structures from homogeneity is $ \sigma_\text{mean}^{\Lambda \text{CDM}} = 0.09 \pm 0.02 $, and the average characteristic size of structures is $ R_\text{mean}^{\Lambda \text{CDM}} = 790 \pm 150 $ Mpc. The maximum size of the detected structure is $ R_j = 1,754 \pm 40 $ Mpc, and the minimum one is $ R_j = 147 \pm 24 $ Mpc. 

\item For the case of the empirical approximation of homogeneity, the average standard deviation of detected structures from homogeneity is $ \sigma_\text{mean}^\text{empiric} = 0.08 \pm 0.01 $, and the average characteristic size of structures is $ R_\text{mean}^\text{empiric} = 640 \pm 140 $ Mpc. The maximum size of the detected structure is $ R_j = 2,000 \pm 80 $ Mpc, and the minimum one is $ R_j = 145 \pm 36 $ Mpc. 

\item We have introduced the selection parameter $w$ to take into account different uncertainty of redshifts $\sigma(z)$. At different values of the parameter $w$, we have obtained the similar results.

\item Our calculations show that the observed cosmic variance of radial fluctuations in the number of galaxies can be also described by the PLCF at all redshifts~\citep{Nabokov2010a, Tekhanovich2016, Shirokov2016}.

\end{itemize}

The method can be also applied in future observation data of the Transient High-Energy Sky and Early Universe Surveyor (THESEUS) space mission project~\citep{amati2018theseus, stratta2018theseus} that together with optical ground-based telescopes, e.g., GTC~\citep{tirado2003study} and BTA~\citep{shirokov2020theseus}, is aimed to explore the unique capabilities of gamma-ray bursts (GRBs) for cosmology and multimessenger astrophysics. 
The observed distribution of galaxies along the line of sight gives information about inhomogeneous distribution of visible matter in the fixed direction in the sky. Statistical analysis of a grid of such fields (towards GRB host-galaxy) will allow one to perform a cosmic tomography of the large-scale distribution of galaxies on largest optically available scales.
The cosmic tomography allows one to constrain precisely cosmology as well as the galaxy structures~\citep{Baryshev2010,Nabokov2010b,Shirokov2016,SokolovJr2018,shirokov2020theseus}.

The codes developed in Python underlying this article are available in the LSA software repository on \emph{www.github.com}, at \href{http://doi.org/10.5281/zenodo.4167356}{DOI: 10.5281/zenodo.4167356}.

\subsection*{The following abbreviations are used in this manuscript:}
\begin{table}[h]
    \begin{tabular}{ll}
LSM & least-squares method\\
LSSU & large-scale structure of Universe\\
SCM & standard cosmological model ($\Lambda$CDM)\\
PLCF & power-law correlation function\\
    \end{tabular}
\end{table}

\subsection*{Authors contribution}
Investigation, M. C., S. S.; 
Methodology, M. N., S. S., V. G.; 
Software, M. N., M. C., A. B.; 
Writing -- original draft, M. N., S. S.; 
Writing -- review \& editing, M. C., S. S., V. G.

\subsection*{Funding}
This research received no external funding.

\subsection*{Conflict of interest}
The authors declare no conflicts of interest.

\acknowledgments{The authors gratefully acknowledge The COSMOS team\footnote{\url{http://cosmos.astro.caltech.edu/page/the-team}} for the free online access to the data. 

We thank the anonymous reviewer for important suggestions that helped us to improve the presentation of our results.
We are grateful to Lovyagin N.~Yu., Gainutdinov R.~I., Parul H.~D., and Baryshev Yu. V. for helpful advices and remarks.

The work was performed as part of the government contract of the SAO RAS approved by the Ministry of Science and Higher Education of the Russian Federation.}


\appendix
\unskip

\bibliographystyle{my_arXiv}
\bibliography{Nikonov.bib}

\section*{Appendix A: Testing of method}
    \label{Appendix A}

The COSMOS2015 catalogue~\citep{Laigle2016} also contains the UltraVISTA DR2~\citep{McCracken2012} catalogue data. The columns \emph{FLAG\_HJMCC} and \emph{FLAG\_DEEP} that have been combined into a single UltraVISTA sample. 
Further, a mutual sampling by the COSMOS and UltraVISTA galaxies have also been done. 
Further we analysed two independent samples: the COSMOS subsample (the optical range) with 184,197 galaxies and the UltraVISTA subsample (the near-IR range) with 40,237 galaxies.

The independence of the samples as well as the coverage of different spectral ranges allows one to conduct their comparative analysis. However, the number of UltraVISTA objects is 5 times less than in the COSMOS so this factor should be taken into account. 
These samples lie in the same direction on the celestial sphere and hence should give a correlation of density fluctuations.

\begin{figure*}  \centering
    
    \includegraphics[width=0.49\linewidth]{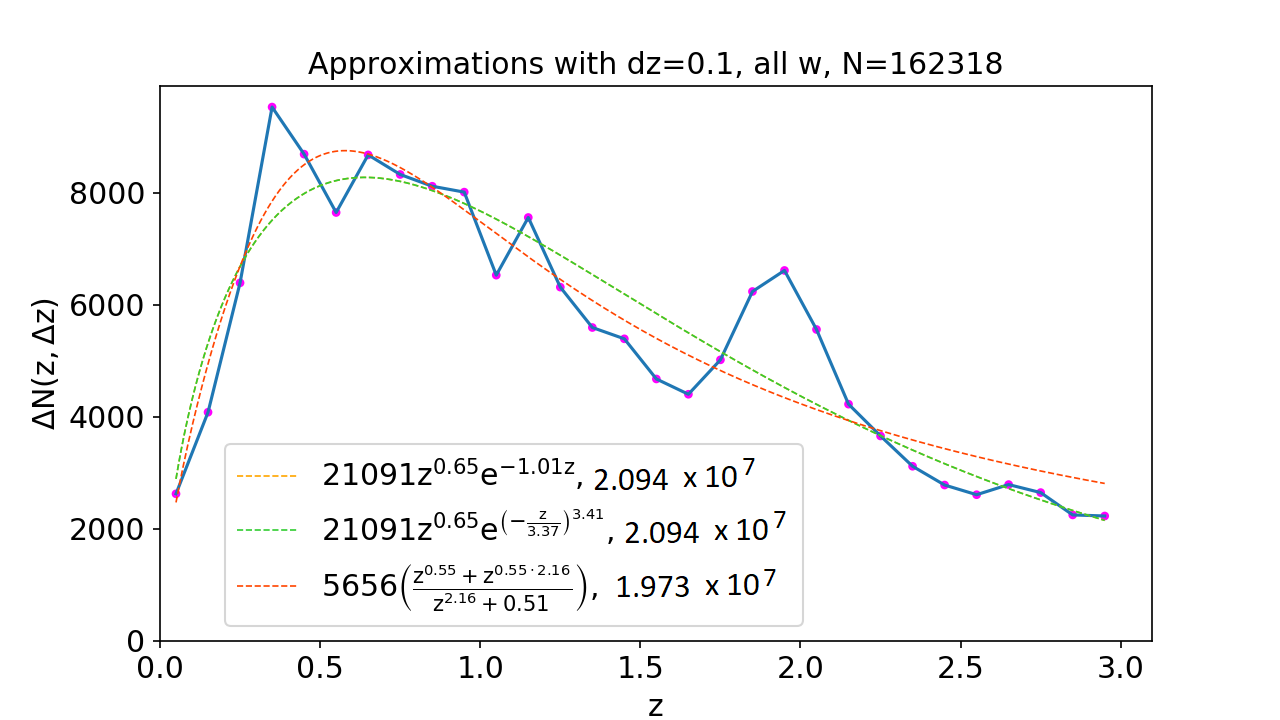}
    \hfill
    \includegraphics[width=0.49\linewidth]{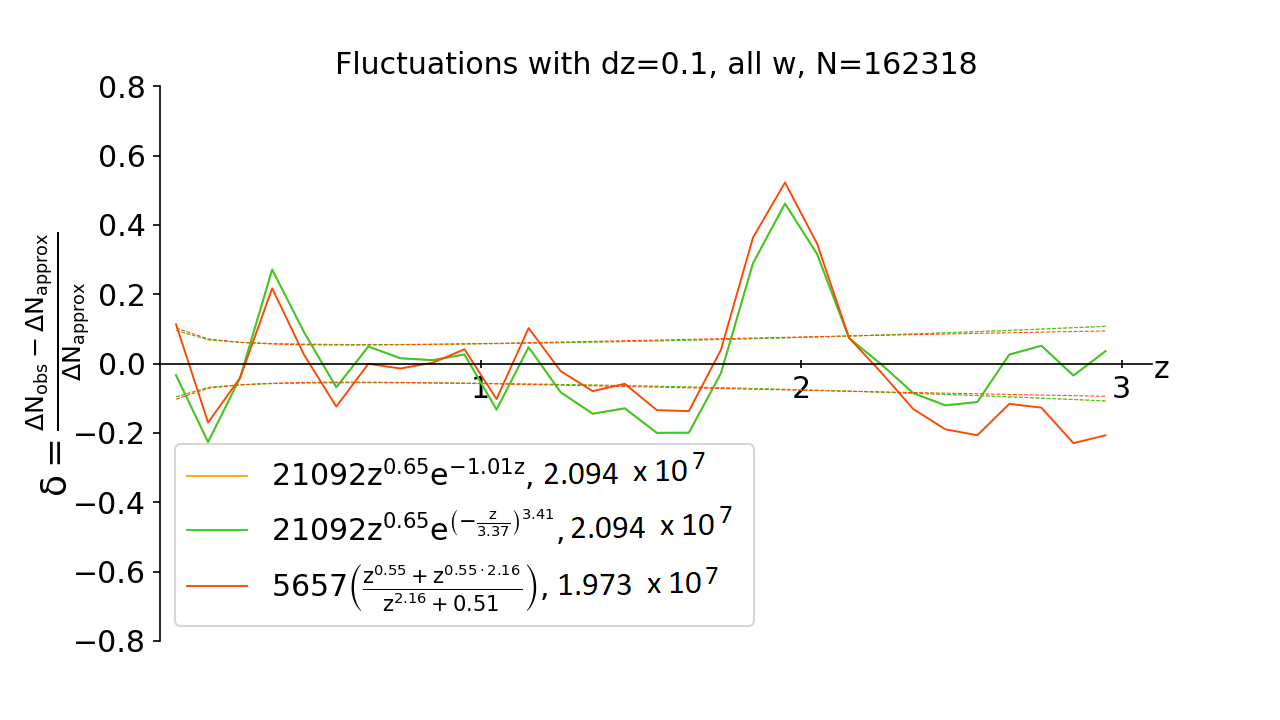}
    \includegraphics[width=0.49\linewidth]{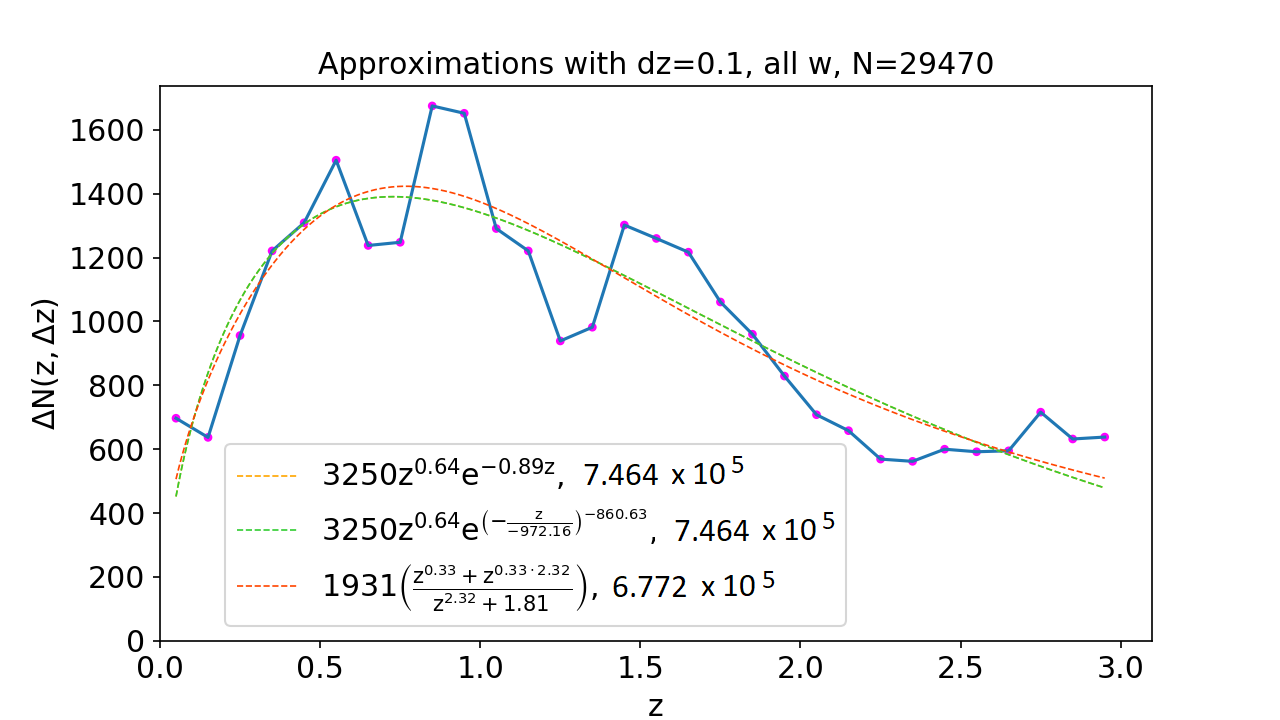}
    \hfill
    \includegraphics[width=0.49\linewidth]{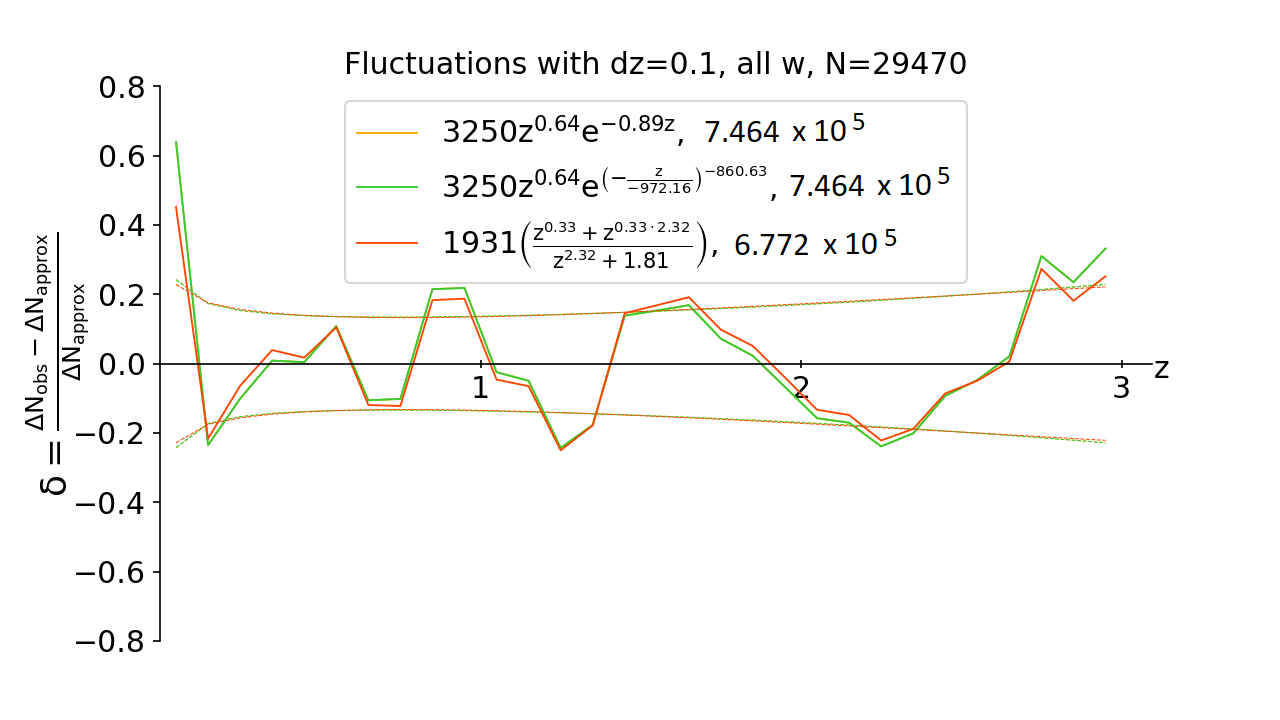}

    \caption{Histograms of the radial distribution of the COSMOS2015 photometric redshifts and corresponding fluctuation patterns for the disjoint 162,318 COSMOS galaxies (top) and 29,470 UltraVISTA galaxies (bottom) with a bin $ \Delta z = 0.1 $ and $ z_\text{max} = 3 $ without sampling by $ \sigma (z) $. The dashed lines indicate the least squares fit (left) and Poisson noise 5$ \sigma $ (right).}
    \label{fig:COSMOS_UltraVISTA_w=all_dz=0.1_zmax=3}
    
\end{figure*}

\begin{figure*}  \centering
    
    \includegraphics[width=0.49\linewidth]{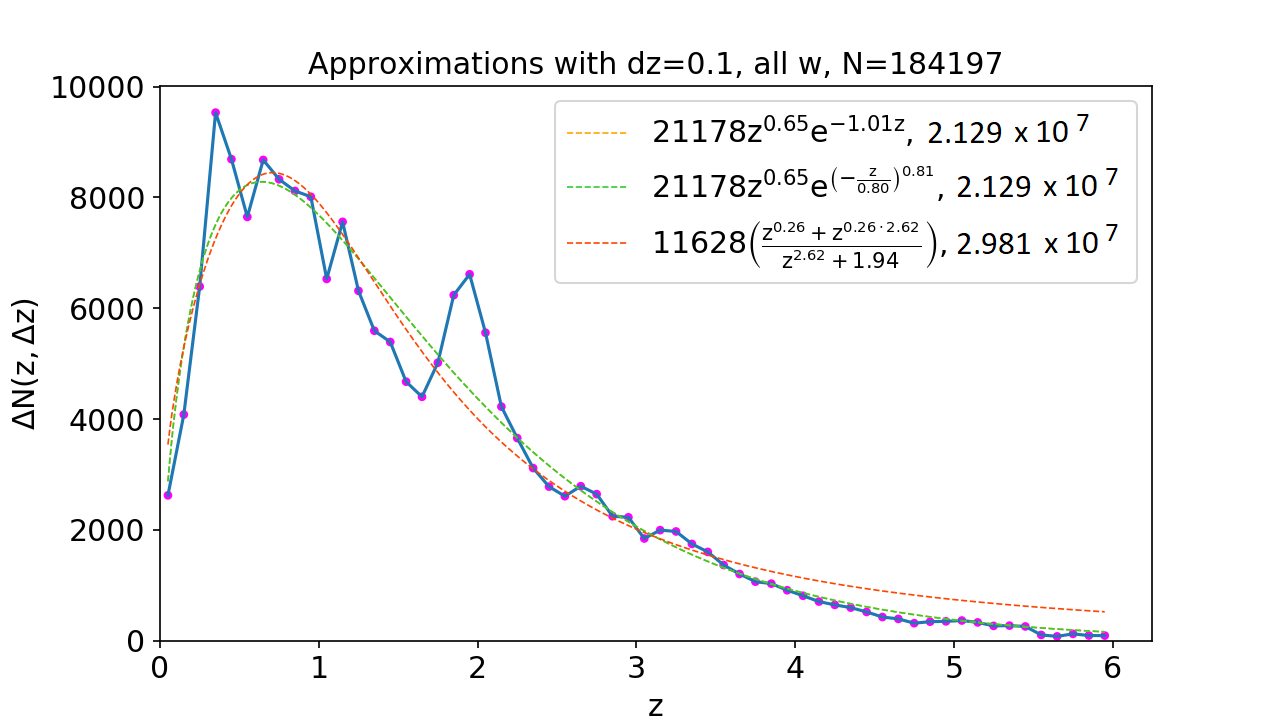}
    \hfill
    \includegraphics[width=0.49\linewidth]{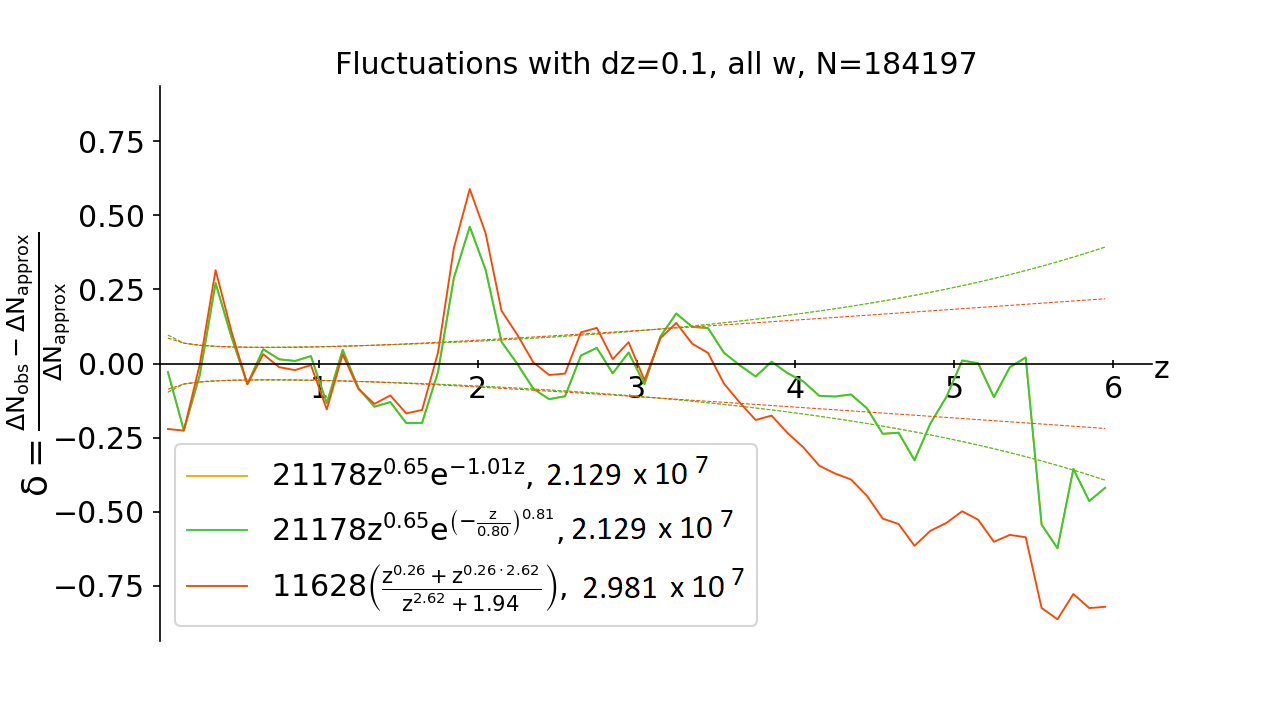}
    \includegraphics[width=0.49\linewidth]{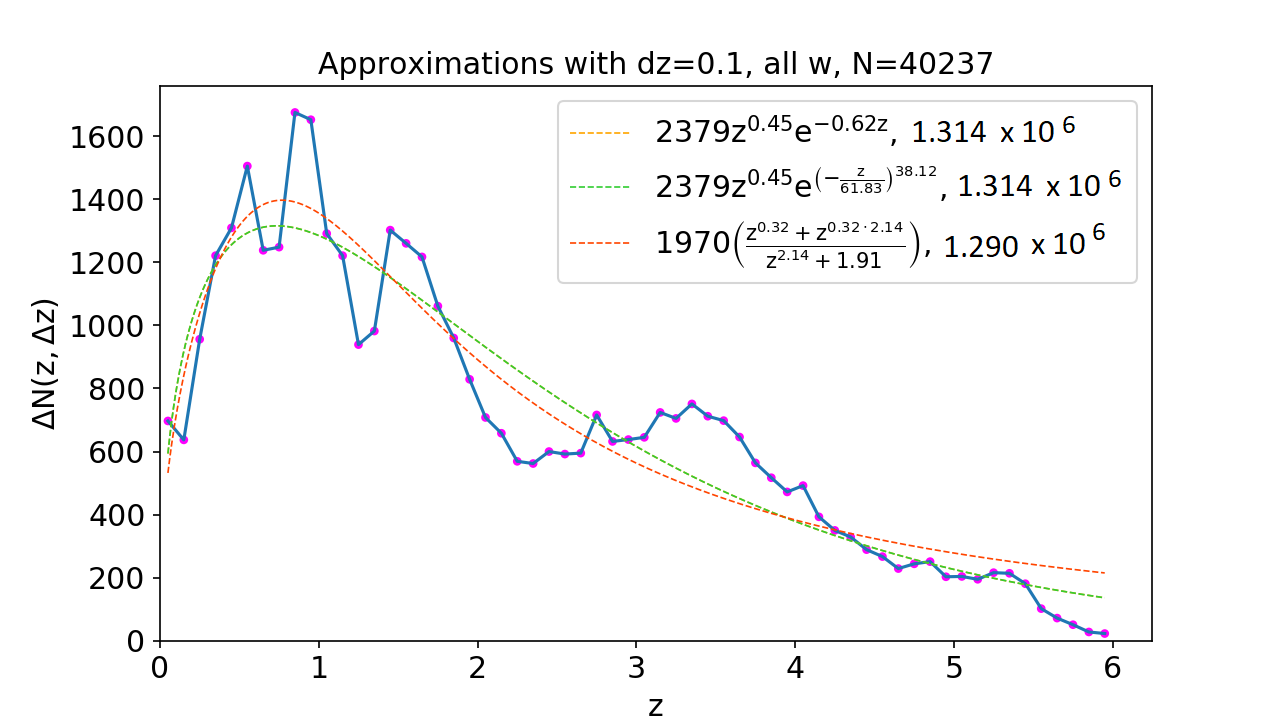}
    \hfill
    \includegraphics[width=0.49\linewidth]{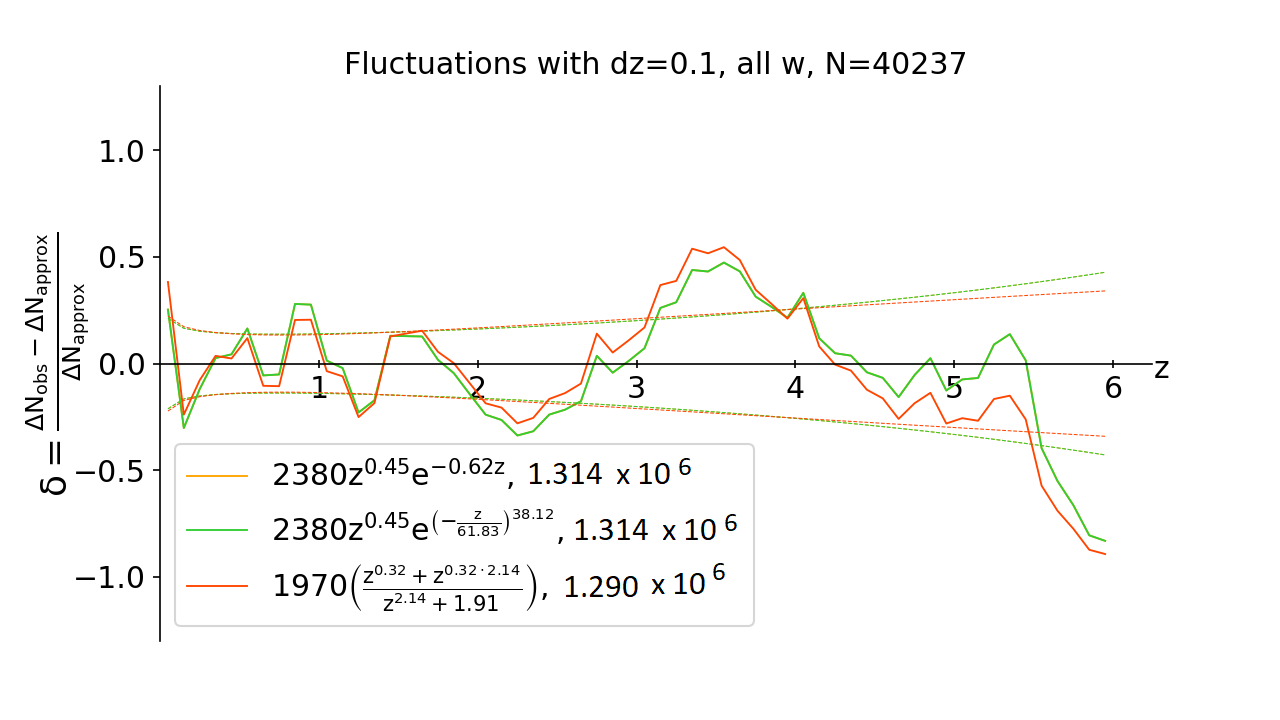}

    \caption {Histograms of radial distribution of the COSMOS2015 photometric redshifts and the corresponding fluctuation patterns for the disjoint COSMOS galaxies (top) and UltraVISTA galaxies (bottom) with a bin $ \Delta z = 0.1 $ and $ z_\text{max} = 6 $ without sampling by $ \sigma (z) $. The dashed lines indicate the least squares fits (left) and Poisson noise 5$ \sigma $ (right).}
    \label{fig:COSMOS_UltraVISTA_w=all_dz=0.1_zmax=6}
    
\end{figure*}

The method was applied with the parameters $ \Delta z = 0.1 $, $ z_\text{max} = (3,6) $ and without $w$-sampling by $ \sigma (z) $.
The corresponding histograms of the radial distribution of galaxies with approximations by the formulas~(\ref{eq:N3(z)}--\ref{eq:N2(z)}) and fluctuation patterns are shown in Figures~\ref{fig:COSMOS_UltraVISTA_w=all_dz=0.1_zmax=3} and~\ref{fig:COSMOS_UltraVISTA_w=all_dz=0.1_zmax=6}. The dotted lines in the fluctuation patterns (right) show the Poisson noise of level 5$ \sigma_\text{P} $. 
As can be seen in the figures, the COSMOS sample shows four distinct (exceeding the level of $ 5 \sigma_\text{P} $) structures with redshifts: $ z = 0.15 \pm 0.10 $ (void), $ z = 0.40 \pm 0.15 $ (cluster), $ z = 1.45 \pm 0.30 $ (void), and $ z = 2.00 \pm 0.25 $ (cluster). The UltraVISTA sample gives a somewhat different picture, although there is a positive correlation with the COSMOS sample at $ z \lesssim 1 $, and shows distinct structures at $ z = 2.25 \pm 0.50 $ (void) and at $ z = 3.65 \pm 0.7 $ (cluster). 

This result can be explained by the fact that in infrared observations the dust attenuation effects are smaller than in optical range. 
Thus, according to the COSMOS2015 catalogue data, it can be concluded that optical observations provide rich statistics, and therefore, are useful for the LSSU analysis at redshifts $ z \lesssim 1 $, while infrared observations have better quality of statistics (in the sense of redshifts) up to $ z \lesssim 6 $.

The structure tables corresponding to Figures~\ref{fig:COSMOS_UltraVISTA_w=all_dz=0.1_zmax=3} and~\ref{fig:COSMOS_UltraVISTA_w=all_dz=0.1_zmax=6} are shown in Tables~\ref{tab:COSMOS_UVISTA_w=all_dz=0.1_zmax=3} for $z_{max}=3$, and~\ref{tab:COSMOS_UVISTA_w=all_dz=0.1_zmax=6} for $z_{max}=6$. 
The parameter $ j $ is the index number of structure, detected by the algorithm. The parameter $ n $ shows the number of bins of a given structure (each structure is a sequence of bins with only positive or only negative fluctuations). The parameter $ \overline{z} $ shows the average redshift value for the bins of each structure, and the errors can be used to restore the centers of the border bins. The parameter $ \Delta r(z) $ gives an estimate of the metric size of the structure (sequence of bins), where the upper and lower limits correspond to the metric size of the border bins. The parameter $ \sigma_\text{P} $ is an estimate of the Poisson noise, which is equal to the ratio of unity to the number of all galaxies in the structure. The parameter $ \sigma_\text{obs} $ shows level of the observed cosmic variance, where the error is the standard deviation from $ \sigma_\text{obs} $ within the structure. We also include columns for comparing observations with the predictions of theoretical models. The parameter $ \sigma_\text{dm} $ shows the theoretical variance of the dark matter density in the $ \Lambda $CDM model frameworks. The next column contains the corresponding galaxy bias $ b_\text{dm} $, with the errors that are proportional to the errors of the $ \sigma_\text{obs} $. 

The next column contains the approximation of the matter density $ \sigma_\text{pl} $, distributed according to the PLCF with the parameters $ (r_0 = 5, \gamma = 1.0) $ (see Tables~3 and~4 in~\citet{Shirokov2016}). At high redshifts ($z>3.6$), linear extrapolation of the data in logarithmic coordinates was used to determine $ \sigma_\text{pl} $. Column $ b_\text{pl} $ contains the corresponding bias parameter, where errors are calculated in the same ways as $ b_\text{dm} $. The bias value $ b = 0 $ indicates that the Poisson noise is higher than $ \sigma_\text{obs} $, and this structure is excluded from the $ b(z) $ function, as well as from the calculation of the average bias value $ b_\text{mean} $.

Last row contains the following average values of corresponding parameters over all structures detected in a sample of galaxies: 
average redshift $z_\text{mean}$, 
average size of structures in Mpc $R_\text{mean}$, 
average Poisson noise level $1\overline{\sigma_\text{P}}$,
average observed cosmic variance $\sigma_\text{mean}$,
average dark matter variance $\overline{\sigma_\text{dm}}$, 
average dark matter bias $\overline{b_\text{dm}}$,
average Peebles correlation function variance $\overline{\sigma_\text{pl}}$
and its bias $\overline{b_\text{pl}}$.
Errors of all the target values correspond to the standard deviation of the corresponding parameters for all structures, except for $ \sigma_\text{mean} $, which is calculated by the formula~(\ref{eq:sigma_mean}).

The catalogue contains an interesting note about redshift columns: ``\emph{a comparison photo-z/spec-z shows that these errors could be underestimated by a factor 0.1 * I -- 0.8 at I > 20 and 1.2 at I < 20}''. Taking this into account one can improve the picture of spatial structures of the catalogue and enhance the correlation of density fluctuations between independent surveys of this field in future studies.

\begin{table*}\centering \scriptsize
    \begin{tabular*}{\textwidth}
{@{\extracolsep{\fill}}ccccccccccc}
           \hline \hline

sample & $j$ & $n$ & $\overline{z}$ & $\Delta r(z)$, Mpc & $\sigma_\text{p}$ &  $\sigma_\text{obs}$ & $\sigma_\text{dm}$ & $b_\text{dm}$ & $\sigma_\text{pl}$ & $b_\text{pl}$ \\ \hline
Only COSMOS	&        1 & 3 & 0.15 $\pm$ 0.10 & $774^{  +203}_{  -183}$ & 0.008 & 0.100 $\pm$ 0.063 & 0.188 & 0.5 $\pm$ 0.3 & 0.199 & 0.5 $\pm$ 0.3 \\ 
$z_\text{max}=3$&        2 & 4 & 0.40 $\pm$ 0.15 & $1012^{  +183}_{  -154}$ & 0.006 & 0.063 $\pm$ 0.078 & 0.140 & 0.5 $\pm$ 0.6 & 0.198 & 0.3 $\pm$ 0.4 \\ 
w=all &       3 & 4 & 0.80 $\pm$ 0.15 & $799^{  +145}_{  -121}$ & 0.006 & 0.025 $\pm$ 0.009 & 0.094 & 0.3 $\pm$ 0.1 & 0.178 & 0.1$\pm$  0.1 \\ 
  &      4 & 3 & 1.05 $\pm$ 0.10 & $459^{  +121}_{  -108}$ & 0.007 & 0.020 $\pm$ 0.057 & 0.070 & 0.3 $\pm$ 0.8 & 0.151 & 0.1 $\pm$ 0.4 \\ 
   &     5 & 7 & 1.45 $\pm$ 0.30 & $1111^{+108}_{  -79}$ & 0.005 & 0.105 $\pm$ 0.034 & 0.073 & 1.4 $\pm$ 0.5 & 0.184 & 0.6$\pm$  0.2 \\ 
    &    6 & 6 & 2.00 $\pm$ 0.25 & $704^{  +79}_{  -62}$ & 0.006 & 0.185 $\pm$ 0.081 & 0.056 & 3.3 $\pm$ 1.4 & 0.168 & 1.1 $\pm$ 0.5 \\ 
     &   7 & 5 & 2.45 $\pm$ 0.20 & $462^{  +62}_{  -53}$ & 0.008 & 0.058 $\pm$ 0.030 & 0.046 & 1.3 $\pm$ 0.6 & 0.154 & 0.4 $\pm$ 0.2 \\ 
      &  8 & 3 & 2.75 $\pm$ 0.10 & $204^{  +53}_{  -49}$ & 0.011 & 0.014 $\pm$ 0.026 & 0.035 & 0.3 $\pm$ 0.4 & 0.125 & 0.1 $\pm$ 0.1 \\ \hline
       means: & & & 1.38 $\pm$ 0.17 & $691^{  +166}_{  -152}$ & 0.007 & 0.071 $\pm$ 0.021 & 0.088 & 1.0 $\pm$ 0.4 & 0.170 & 0.4 $\pm$ 0.1 \\ \hline \hline

    sample & $j$ & $n$ & $\overline{z}$ & $\Delta r(z)$, Mpc & $\sigma_\text{p}$ &  $\sigma_\text{obs}$ & $\sigma_\text{dm}$ & $b_\text{dm}$ & $\sigma_\text{pl}$ & $b_\text{pl}$ \\ \hline
Only UltraVISTA	&        1 & 2 & 0.10 $\pm$ 0.05 & $397^{  +203}_{  -193}$ & 0.028 & 0.202 $\pm$ 0.438 & 0.157 & 1.3 $\pm$ 2.8 & 0.155 & 1.3 $\pm$ 2.8 \\ 
$z_\text{max}=3$&        2 & 3 & 0.25 $\pm$ 0.10 & $734^{  +193}_{  -173}$ & 0.018 & 0.109 $\pm$ 0.071 & 0.155 & 0.7 $\pm$ 0.5 & 0.187 & 0.6 $\pm$ 0.4 \\ 
w=all &       3 & 4 & 0.50 $\pm$ 0.15 & $955^{  +173}_{  -145}$ & 0.014 & 0.004 $\pm$ 0.044 & 0.124 & 0.0 $\pm$ 0.0 & 0.192 & 0.0 $\pm$ 0.0 \\ 
  &      4 & 4 & 0.90 $\pm$ 0.15 & $753^{  +137}_{  -114}$ & 0.014 & 0.077 $\pm$ 0.082 & 0.087 & 0.9 $\pm$ 0.9 & 0.174 & 0.4 $\pm$ 0.5 \\ 
   &     5 & 5 & 1.25 $\pm$ 0.20 & $823^{  +114}_{  -92}$ & 0.013 & 0.071 $\pm$ 0.066 & 0.075 & 0.9 $\pm$ 0.9 & 0.176 & 0.4 $\pm$ 0.4 \\ 
    &    6 & 5 & 1.65 $\pm$ 0.20 & $666^{  +92}_{  -75}$ & 0.014 & 0.111 $\pm$ 0.028 & 0.062 & 1.8 $\pm$ 0.4 & 0.167 & 0.7 $\pm$ 0.2 \\ 
     &   7 & 9 & 2.25 $\pm$ 0.40 & $1012^{  +75}_{  -53}$ & 0.012 & 0.104 $\pm$ 0.031 & 0.055 & 1.9 $\pm$ 0.6 & 0.175 & 0.6 $\pm$ 0.2 \\ \hline
      means: & & & 0.99 $\pm$ 0.18 & $763^{  +170}_{  -151}$ & 0.016 & 0.097 $\pm$ 0.023 & 0.102 & 1.2 $\pm$ 0.3 & 0.175 & 0.7 $\pm$ 0.2 \\ \hline 
      
 \end{tabular*}
 
    \caption{
    Tables of structures for disjoint the COSMOS (top) and UltraVISTA (bottom) samples from the COSMOS2015 catalogue for a bin $\Delta z = 0.1$ with $z_\text{max}=3$ and without $\sigma_z$ selection. The approximation is by empirical formula (\ref{eq:N2(z)}).
    In the last string there are the means of corresponding values (by all structures): 
    mean of redshifts $z_\text{mean}$, 
    mean of structure sizes in Mpc $R_\text{mean}$, 
    mean of Poisson noise level $1\overline{\sigma_\text{P}}$,
    mean of observed cosmic variance $\sigma_\text{mean}$,
    mean of dark matter variance $\overline{\sigma_\text{dm}}$, 
    mean of dark matter bias $\overline{b_\text{dm}}$,
    mean of Peebles correlation function variance $\overline{\sigma_\text{pl}}$
    and its bias $\overline{b_\text{pl}}$ (see details in the text).
    }
    \label{tab:COSMOS_UVISTA_w=all_dz=0.1_zmax=3}
\end{table*}

\begin{table*}\centering \scriptsize
    \begin{tabular*}{\textwidth}
{@{\extracolsep{\fill}}ccccccccccc}
       \hline \hline    
sample & $j$ & $n$ & $\overline{z}$ & $\Delta r(z)$, Mpc & $\sigma_\text{p}$ &  $\sigma_\text{obs}$ & $\sigma_\text{dm}$ & $b_\text{dm}$ & $\sigma_\text{pl}$ & $b_\text{pl}$ \\ \hline
Only COSMOS	&        1 & 3 & 0.15 $\pm$ 0.10 & $774^{  +203}_{  -183}$ & 0.008 & 0.098 $\pm$ 0.064 & 0.188 & 0.5 $\pm$ 0.3 & 0.199 & 0.5 $\pm$ 0.3 \\ 
$z_\text{max}=6$&        2 & 4 & 0.40 $\pm$ 0.15 & $1012^{  +183}_{  -154}$ & 0.006 & 0.063 $\pm$ 0.078 & 0.140 & 0.5 $\pm$  0.6 & 0.198 & 0.3 $\pm$ 0.4 \\ 
w=all &       3 & 4 & 0.80 $\pm$ 0.15 & $799^{  +145}_{  -121}$ & 0.006 & 0.024 $\pm$ 0.009 & 0.094 & 0.3 $\pm$ 0.1 & 0.178 & 0.1 $\pm$ 0.1 \\ 
  &      4 & 3 & 1.05 $\pm$ 0.10 & $459^{  +121}_{  -108}$ & 0.007 & 0.021 $\pm$ 0.057 & 0.070 & 0.3 $\pm$ 0.8 & 0.151 & 0.1 $\pm$ 0.4 \\ 
   &     5 & 7 & 1.45 $\pm$ 0.30 & $1111^{  +108}_{  -79}$ & 0.005 & 0.106 $\pm$ 0.034 & 0.073 & 1.5 $\pm$ 0.5 & 0.184 & 0.6$\pm$  0.2 \\ 
    &    6 & 6 & 2.00 $\pm$ 0.25 & $704^{  +79}_{  -62}$ & 0.006 & 0.185 $\pm$ 0.081 & 0.056 & 3.3 $\pm$ 1.4 & 0.168 & 1.1 $\pm$ 0.5 \\ 
     &   7 & 5 & 2.45 $\pm$ 0.20 & $462^{  +62}_{  -53}$ & 0.008 & 0.058 $\pm$ 0.030 & 0.046 & 1.2 $\pm$ 0.6 & 0.154 & 0.4 $\pm$ 0.2 \\ 
      &  8 & 3 & 2.75 $\pm$ 0.10 & $204^{  +53}_{  -49}$ & 0.012 & 0.016 $\pm$ 0.025 & 0.035 & 0.3 $\pm$ 0.5 & 0.125 & 0.1 $\pm$ 0.1 \\ 
       & 9 & 7 & 3.35 $\pm$ 0.30 & $497^{  +45}_{  -37}$ & 0.010 & 0.066 $\pm$ 0.032 & 0.039 & 1.7 $\pm$ 0.8 & 0.155 & 0.4 $\pm$ 0.2 \\ 
&        10 & 3 & 3.75 $\pm$ 0.10 & $145^{  +37}_{  -35}$ & 0.017 & 0.014 $\pm$ 0.015 & 0.028 & 0.0 $\pm$ 0.0 & 0.118 & 0.0 $\pm$ 0.0 \\ 
 &       11 & 13 & 4.45 $\pm$ 0.60 & $717^{  +35}_{  -25}$ & 0.011 & 0.128 $\pm$ 0.028 & 0.034 & 3.7 $\pm$ 0.8 & 0.159 & 0.8 $\pm$ 0.2 \\
  &      12 & 3 & 5.25 $\pm$ 0.10 & $96^{  +24}_{  -23}$ & 0.033 & 0.041 $\pm$ 0.036 & 0.022 & 1.2 $\pm$ 1.0 & 0.112 & 0.2 $\pm$ 0.2 \\ \hline
   means: & & & 2.32 $\pm$ 0.20 & $582^{  +134}_{  -124}$ & 0.011 & 0.068 $\pm$ 0.015 & 0.069 & 1.3 $\pm$ 0.4 & 0.158 & 0.4 $\pm$ 0.1 \\ \hline \hline

sample & $j$ & $n$ & $\overline{z}$ & $\Delta r(z)$, Mpc & $\sigma_\text{p}$ &  $\sigma_\text{obs}$ & $\sigma_\text{dm}$ & $b_\text{dm}$ & $\sigma_\text{pl}$ & $b_\text{pl}$ \\ \hline
Only UltraVISTA	&  1 & 3 & 0.25 $\pm$ 0.10 & $734^{  +193}_{  -173}$ & 0.018 & 0.132 $\pm$ 0.095 & 0.155 & 0.9 $\pm$ 0.6 & 0.187 & 0.7 $\pm$ 0.5 \\ 
$z_\text{max}=6$ & 2 & 4 & 0.50 $\pm$ 0.15 & $955^{  +173}_{  -145}$ & 0.014 & 0.045 $\pm$ 0.045 & 0.124 & 0.3 $\pm$ 0.4 & 0.192 & 0.2 $\pm$ 0.2 \\ 
w=all    &    3 & 4 & 0.90 $\pm$ 0.15 & $753^{  +137}_{  -114}$ & 0.014 & 0.130 $\pm$ 0.087 & 0.087 & 1.5 $\pm$ 1.0 & 0.174 & 0.7 $\pm$ 0.5 \\ 
     &   4 & 5 & 1.25 $\pm$ 0.20 & $823^{  +114}_{ -92}$ & 0.013 & 0.056 $\pm$ 0.065 & 0.075 & 0.7 $\pm$ 0.8 & 0.176 & 0.3 $\pm$ 0.4 \\ 
      &  5 & 4 & 1.60 $\pm$ 0.15 & $511^{  +92}_{  -79}$ & 0.015 & 0.101 $\pm$ 0.028 & 0.059 & 1.7 $\pm$ 0.5 & 0.158 & 0.6 $\pm$ 0.2 \\ 
       & 6 & 11 & 2.25 $\pm$ 0.50 & $1271^{  +79}_{  -51}$ & 0.010 & 0.175 $\pm$ 0.039 & 0.057 & 3.1 $\pm$ 0.7 & 0.179 & 1.0 $\pm$ 0.2 \\ 
&        7 & 3 & 2.85 $\pm$ 0.10 & $197^{  +51}_{  -47}$ & 0.022 & 0.002 $\pm$ 0.023 & 0.035 & 0.0 $\pm$ 0.0 & 0.124 & 0.0 $\pm$ 0.0 \\ 
 &       8 & 15 & 3.65 $\pm$ 0.70 & $1063^{  +47}_{  -30}$ & 0.012 & 0.249 $\pm$ 0.041 & 0.040 & 6.2 $\pm$ 1.0 & 0.166 & 1.5 $\pm$ 0.3 \\ 
  &      9 & 6 & 4.60 $\pm$ 0.25 & $285^{  +30}_{  -26}$ & 0.024 & 0.043 $\pm$ 0.029 & 0.030 & 1.2 $\pm$ 0.8 & 0.143 & 0.3 $\pm$ 0.2 \\ 
   &     10 & 4 & 5.00 $\pm$ 0.15 & $154^{  +26}_{  -24}$ & 0.033 & 0.061 $\pm$ 0.032 & 0.025 & 2.0 $\pm$ 1.0 & 0.127 & 0.4 $\pm$ 0.2 \\ 
    &    11 & 4 & 5.30 $\pm$ 0.15 & $143^{  +24}_{  -23}$ & 0.036 & 0.044 $\pm$ 0.045 & 0.024 & 1.0 $\pm$ 1.1 & 0.126 & 0.2 $\pm$ 0.2 \\ \hline
     means: & & & 2.56 $\pm$ 0.24 & $626^{  +150}_{  -141}$ & 0.019 & 0.094 $\pm$ 0.023 & 0.065 & 1.9 $\pm$ 0.5 & 0.159 & 0.6 $\pm$ 0.1 \\ \hline 

    \end{tabular*}
    
    \caption{
    Tables of structures for disjoint the COSMOS (top) and UltraVISTA (bottom) samples from the COSMOS2015 catalogue for a bin $\Delta z = 0.1$ with $z_\text{max}=6$ and without $\sigma_z$ selection. The approximation is by empirical formula (\ref{eq:N2(z)}).
    In the last string there are the means of corresponding values (by all structures): 
    mean of redshifts $z_\text{mean}$, 
    mean of structure sizes in Mpc $R_\text{mean}$, 
    mean of Poisson noise level $1\overline{\sigma_\text{P}}$,
    mean of observed cosmic variance $\sigma_\text{mean}$,
    mean of dark matter variance $\overline{\sigma_\text{dm}}$, 
    mean of dark matter bias $\overline{b_\text{dm}}$,
    mean of Peebles correlation function variance $\overline{\sigma_\text{pl}}$
    and its bias $\overline{b_\text{pl}}$ (see details in the text).
    }
    \label{tab:COSMOS_UVISTA_w=all_dz=0.1_zmax=6}
\end{table*}

\section*{Appendix B: The metric space}

\begin{figure}  \centering
    
    \includegraphics[width=\linewidth]{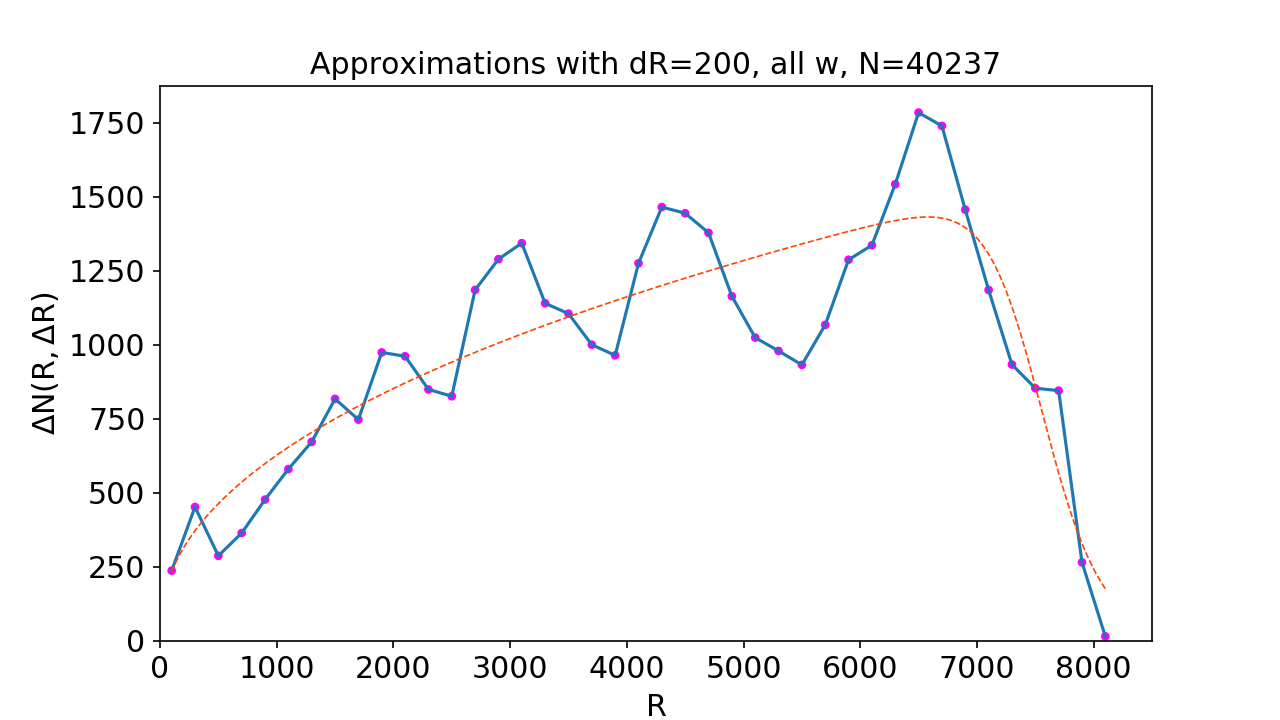}
    
    \caption{Histogram of radial distribution of the Deep-UltraVISTA~\citep{Laigle2016} photometric redshifts recalculated to metric distances in comoving space with a bin $ \Delta R = 200 $ Mpc without sampling by $ \sigma (z) $. The dashed line indicates the least squares fit by the theoretical approximation of homogeneity (according to the SCM) by Eq.~(\ref{eq:N3(z)}).}
    \label{fig:dR}
    
\end{figure}

Figure~\ref{fig:dR} shows an example of the histogram of radial distribution of galaxies in the metric space (in comoving coordinate system) calculated in the SCM frameworks with parameters ($ H_0 = 70 $, $ \Omega_\text{v} = 0.7 $ ). The transition from catalogs of redshifts to catalogs of metric distances on a grid of models will be possible with more complete statistics of galaxies obtained by narrow-angle deep surveys like the COSMOS. In the future, this approach could impose new restrictions on cosmological parameters obtained by the cosmic tomography (e.g.,~\citet{Baryshev2010,Nabokov2010b,shirokov2020theseus}). 

\end{document}